\documentclass{sigchi}
\usepackage{balance}
\usepackage{graphicx}
\usepackage{times}
\usepackage[hyphens]{url}

\usepackage{adjustbox}
\usepackage{eurosym}

\makeatletter
\def\url@leostyle{%
  \@ifundefined{selectfont}{\def\UrlFont{\sf}}{\def\UrlFont{\small\bf\ttfamily}}}
\makeatother
\makeatletter
\def\@copyrightspace{\relax}
\makeatother

\def\pprw{8.5in}
\def\pprh{11in}

\usepackage{hyperref}
\hypersetup{
colorlinks,
citecolor=blue,
filecolor=blue,
linkcolor=blue,
urlcolor=blue,
breaklinks=true,
}

\usepackage[hyphenbreaks]{breakurl}
\renewcommand{\footnotesize}{\scriptsize}

\begin{document}
\title{Facebook Use of Sensitive Data for Advertising in Europe}

\numberofauthors{3}
\author{
  \alignauthor Jos\'{e} Gonz\'{a}lez Caba\~{n}as\\
    \affaddr{Universidad Carlos III de Madrid}\\
    \email{jgcabana@it.uc3m.es}
  \alignauthor \'{A}ngel Cuevas\\
    \affaddr{Universidad Carlos III de Madrid}\\
    \email{acrumin@it.uc3m.es}
  \alignauthor Rub\'{e}n Cuevas\\
    \affaddr{Universidad Carlos III de Madrid}\\
    \email{rcuevas@it.uc3m.es}
}
\maketitle

\begin{abstract}

The upcoming European General Data Protection Regulation (GDPR) prohibits the processing and exploitation of some categories of personal data (health, political orientation, sexual preferences, religious beliefs, ethnic origin, etc.) due to the obvious privacy risks that may be derived from a malicious use of such type of information. These categories are referred to as sensitive personal data. Facebook has been recently fined \euro1.2M in Spain for collecting, storing and processing sensitive personal data for advertising purposes. This paper quantifies the portion of Facebook users in the European Union (EU) who are labeled with interests linked to sensitive personal data. The results of our study reveal that Facebook labels 73\% EU  users with sensitive interests. This corresponds to 40\% of the overall EU population. We also estimate that a malicious third-party could unveil the identity of Facebook users that have been assigned a sensitive interest at a cost as low as \euro0.015 per user. Finally, we propose and implement a web browser extension to inform Facebook users of the sensitive interests Facebook has assigned them.    
\end{abstract}


\section{Introduction}
The citizens of the European Union (EU) show serious concerns regarding the management of personal information by online services. The 2015 Eurobarometer about data protection \cite{Eurobarometer} reveals that: 63\% of EU citizens do not trust online businesses, more than half do not like providing personal information in return for free services, and 53\% do not like that Internet companies use their personal information in tailored advertising. The EU reacted to citizens' concerns with the approval of the General Data Protection Regulation (GDPR) \cite{GDPR}, which defines a new regulatory framework in the management of personal information. EU member states were given a period of two years to transpose it into their national legislation, until May 2018.

The GDPR (as well as many other EU national data protection laws) defines some categories of personal data as very sensitive and imposes the prohibition to process them with very few exceptions (e.g., the user provides explicit consent to process that data for some specific purpose). These categories of data are indistinguishably referred to as \textit{``Specially Protected Data"}, \textit{``Special Categories of Personal Data"} or \textit{``Sensitive Data"}. In particular, the GDPR defines as sensitive personal data:  \textit{``data revealing racial or ethnic origin, political opinions, religious or philosophical beliefs, or trade union membership, and the processing of genetic data, biometric data for the purpose of uniquely identifying a natural person, data concerning health or data concerning a natural person's sex life or sexual orientation"}.  

Due to the obvious legal, ethical and privacy implications of processing sensitive personal data, it is important to unveil whether online services are commercially exploiting such sensitive information. In case it is happening, it is also essential to measure the size of the problem. That means, the portion of users/citizens who may be affected by the exploitation of their sensitive personal data. In this paper, we address these crucial questions focusing in \emph{online advertising}, which represents the most important source of revenue for most online services. In particular, we consider Facebook (FB), the second most important advertising platform (just after Google) in this market.


\begin{figure}[t]
\centering
\includegraphics[width=1\columnwidth]{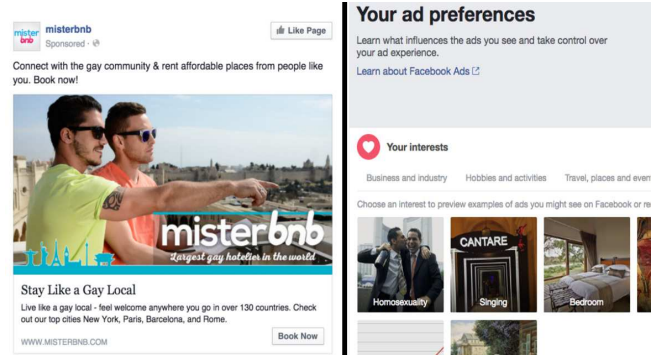}\hfill
\caption{Snapshot of an ad received by one of the authors papers \& ad preference list showing that FB have inferred this person was interested in \textit{Homosexuality}.}
\label{fig:ad1}
\end{figure}

This research focuses in FB due to an episode that happened to one of the authors of this paper. He received the ad shown on the left side of Figure \ref{fig:ad1}. The referred ad stated: \textit{``Connect with the gay community \& rent affordable places from people like you. Book Now"}. The company running the ad campaign is an online service targeting the gay community. 
FB users are labeled with the so-called \emph{ad preferences}, which represent potential interests of the users. FB assigns a user different ad preferences based on her online activity within this social network and in third-party websites tracked by FB.  Advertisers running ad campaigns target groups of users that have been assigned a particular ad preference (e.g., target FB users interested in \textit{``Starbucks"}). The referred author found that \textit{``Homosexuality"} was one of the ad preferences FB had assigned him (see Figure \ref{fig:ad1} right side). He had neither explicitly defined his sexual orientation in any FB setting, nor granted explicit permission to FB to be targeted based on his sexual orientation. 

This episode illustrates that FB is actually processing sensitive personal information, which is prohibited in the upcoming EU GDPR and also in some national data protection regulations in Europe. This claim has been recently confirmed when the Spanish Data Protection Agency (DPA) fined FB with \euro1.2M for violating the Spanish data protection regulation \cite{AEPD}. The Spanish DPA argued that FB \textit{``collects, stores and uses data, including specially protected data, for advertising purposes without obtaining consent."} 

Motivated by these events and the upcoming enforcement of the GDPR in the European Union, the main goal of this paper is \textit{quantifying the portion of EU citizens and FB users that have been assigned ad preferences linked to sensitive personal data}. 

 
To achieve our goal we analyze more than 5.5M ad preferences (126K unique) assigned to more than 4.5K FB users who have installed the Data Valuation Tool for Facebook Users (FDVT) browser extension \cite{FDVT}. The reason for using ad preferences assigned to FDVT users is that we can prove the ad preferences considered in our study have been indeed assigned to real users. 


The first contribution of this paper is a methodology that combines natural language processing techniques and manual classification conducted by 12 panelists to obtain those ad preferences in our dataset linked to sensitive personal data. These are ad preferences that may be used to reveal: ethnic or racial origin, political opinions, religious beliefs, health information or sexual orientation. For instance, the ad preferences \textit{``Homosexuality"} and \textit{``Communism"} may reveal the sexual orientation and the political preference of a user, respectively.

Once we have identified the list of sensitive ad preferences, we use it to query the FB Ads Manager in order to obtain the number of FB users and citizens exposed to sensitive ad preferences in the whole EU as well as in each one of its member states. This quantification is our second contribution, which accomplishes the main goal of the paper. 

Finally, after illustrating privacy and ethics risks derived from the exploitation of FB sensitive ad preferences, we present an extension of the FDVT that informs users of the sensitive ad preferences FB has assigned them. This is the last contribution of this paper.  

Our research leads to the following main insights:

\textbf{We have identified 2092 (1.66\%) sensitive ad preferences out of the 126k present in our dataset.}

\textbf{FB assigns in average 16 sensitive ad preferences to FDVT users.}

\textbf{More than 73\% EU FB users, which corresponds to 40\% EU citizens, are labeled with at least one of the Top 500 (i.e., most popular) sensitive ad preferences from our dataset.}

\textbf{Women show a significantly higher exposure than men to sensitive ad preferences. Similarly, The Early Adulthood group (20-39 years old) is the one showing more exposure.}

\textbf{We perform a ball-park estimation that suggests that unveiling the identity of FB users labeled with sensitive ad preferences may be as cheap as \euro0.015 per user.}






\section{Background}
\subsection{Facebook Ads Manager}
\label{subsec:ads_manager}
Advertisers configure their ads campaigns through the Facebook (FB) Ads Manager\footnote{\url{https://www.facebook.com/ads/manager}}. It allows advertisers to define the audience (i.e., user profile) they want to target with their advertising campaigns. It can be accessed through either a dashboard or an API. The FB Ads Manager offers advertisers a wide range of configuration parameters such as (but not limited to): \emph{location} (country, region, city, zip code, etc.), \emph{demographic parameters} (gender, age, language, etc.), \emph{behaviors} (mobile device, OS and/or web browser used, traveling frequency, etc.), \emph{interests} (sports, food, cars, beauty, etc.).

The \emph{interest} parameter is the most relevant for our work. It includes hundreds of thousands of possibilities capturing users' interest of any type. These interests are organized in a hierarchical structure with several levels. The first level is formed by 14 categories\footnote{Business and industry, Education, Family and relationships, Fitness and wellness, Food and drink, Hobbies and activities, Lifestyle and culture, News and entertainment, People, Shopping and fashion, Sports and outdoors, Technology, Travel places and events, Empty.}. In addition to the interests included in this hierarchy, the FB Ads Manager offers the \textit{Detailed Targeting} search bar where users can type any free text and it suggests interests linked to such text. In this paper, we will leverage the \emph{interest} parameter to identify potential sensitive interests. 

Advertisers can configure their target audiences based on any combination of the described parameters. An example of an audience could be \textit{``Users living in Italy, ranging between 30 and 40 years old, male and interested in Fast Food"}.

Finally, the FB Ads Manager provides detailed information about the configured audience 
from which the most relevant parameter for our paper is the \emph{Potential Reach} that reports the number of registered FB users matching the defined audience.




\subsection{Facebook ad preferences}
\label{subsec:ads_preferences}
FB assigns to each user a set of ad preferences, i.e., a set of interests, derived from the data and activity of the user in FB and external websites, apps and online services where FB is present. These ad preferences are indeed the interests offered to advertisers in the FB Ads Manager to configure their audiences\footnote{Note that given that interests and ad preferences refer to the same thing, we use these two terms indistinguishably in the rest of the paper.}. Therefore, if a user is assigned \textit{``Watches"} within her list of ad preferences, she will be a potential target of any FB advertising campaign configured to reach users interested in watches.

Any user can access and edit (add or remove) her ad preferences\footnote{Access and edit ad preference list: \url{https://facebook.com/ads/preferences/edit}}, but very few users are aware of this option. When a user positions the mouse on a specific ad preference item, a pop-up text indicates why the user has been assigned this ad preference. 
By examining 5.5M ad preferences assigned to FDVT users (see Subsection \nameref{subsec:FDVT}), we have found 6 reasons for the assignment of ad preferences:
$(i)$ \textit{This is a preference you added}, $(ii)$ \textit{You have this preference because we think it may be relevant to you based on what you do on Facebook, such as pages you've liked or ads you've clicked}, $(iii)$ \textit{You have this preference because you clicked on an ad related to...}, $(iv)$ \textit{You have this preference because you installed the app...}, $(v)$ \textit{You have this preference because you liked a Page related to...}, $(vi)$ \textit{You have this preference because of comments, posts, shares or reactions you made related to...} 





\subsection{FDVT}
\label{subsec:FDVT}
The \emph{Data Valuation Tool for Facebook Users (FDVT)} \cite{FDVT} is a web browser extension currently available for Google Chrome\footnote{\url{https://chrome.google.com/webstore/detail/fdvt-social-network-data/blednbbpnnambjaefhlocghajeohlhmh}} and Mozilla Firefox\footnote{\url{https://addons.mozilla.org/firefox/addon/fdvt}}. It provides FB users with a real-time estimation of the revenue they are generating for Facebook according to their profile and the number of ads they visualize and click during a Facebook session. More than 6K users have installed the FDVT between its public release in October 2016 and February 2018.
The FDVT collects (among other data) the ad preferences FB assigns to the user. We will leverage this information to identify sensitive ad preferences assigned to users that have installed the FDVT.

\section{Legal Considerations}
\subsection{General Data Protection Regulation}
\label{sec:legal}
The forthcoming EU General Data Protection Regulation (GDPR) \cite{GDPR} will enter into force in May 2018 and will be the reference data protection regulation in all 28 EU countries. 
The GDPR includes an article that regulates the use of \textit{Sensitive Personal Data}. Article 9 is entitled \textit{``Processing of special categories of personal data"} and literally states in its first paragraph: \textit{``Processing of personal data revealing racial or ethnic origin, political opinions, religious or philosophical beliefs, or trade union membership, and the processing of genetic data, biometric data for the purpose of uniquely identifying a natural person, data concerning health or data concerning a natural person's sex life or sexual orientation shall be prohibited"}.

After enunciating the prohibition of processing any of the items listed in the paragraph, the GDPR introduces in its second paragraph 10 exceptions in which the paragraph 1 of the article shall not apply. 
None of these exceptions apply to the processing and exploitation of sensitive personal data through ad preferences in the case of FB. 

Following we list the exceptions included in GDPR Article 9 that allow processing sensitive information and discuss for each of them the reason why they do not apply to the case of FB ad preferences. In the exceptions text, the term data subject refers to users in the context of FB and the term data controller refers to FB itself.

(a) \textit{``the data subject has given explicit consent to the processing of those personal data for one or more specified purposes, except where Union or Member State law provide that the prohibition referred to in paragraph 1 may not be lifted by the data subject".} 

In most cases, ad preferences are inferred out of the activity of the user in FB (e.g., like a page, click on an ad, etc.). However, none of these activities imply an explicit consent of the user to be assigned ad preferences related to sensitive personal data, which is exploited for advertising purposes. For instance, in the example described in the introduction involving one of the paper authors, this person did not provide any explicit consent that allows FB processing data related to his sexual orientation. In addition, there are no reasons why the user could not lift explicit consent to FB to process her sensitive personal data for advertising purposes. As we described in Section \nameref{subsec:fb_ts}, by accepting the FB Terms of Service, users grant permission to process and store personal data, but there is no reference to sensitive personal data. FB should implement an explicit and transparent process to obtain explicit permission to use sensitive personal data, which is not the case nowadays.

(b) \textit{``processing is necessary for the purposes of carrying out the obligations and exercising specific rights of the controller or of the data subject in the field of employment and social security and social protection law in so far as it is authorised by Union or Member State law or a collective agreement pursuant to Member State law providing for appropriate safeguards for the fundamental rights and the interests of the data subject".}

This exception does not apply to FB since processing sensitive personal data for ad preferences is not related to employment or social security field. Similarly, FB activity is not related to the protection of fundamental rights and interests of FB users.

(c) \textit{``processing is necessary to protect the vital interests of the data subject or of another natural person where the data subject is physically or legally incapable of giving consent".} 

FB has nothing to do with the vital interests of FB users. In addition, the vast majority of FB users are physically and legally capable of giving explicit consent. Therefore, this exception does not apply to FB.

(d) \textit{``processing is carried out in the course of its legitimate activities with appropriate safeguards by a foundation, association or any other not-for-profit body with a political, philosophical, religious or trade union aim and on condition that the processing relates solely to the members or to former members of the body or to persons who have regular contact with it in connection with its purposes and that the personal data are not disclosed outside that body without the consent of the data subject."} 

FB is a private profit company that does not match the organizations described in this exception. Therefore, this exception does not apply to the case of FB.

(e) \textit{``processing relates to personal data which are manifestly made public by the data subject."} 

In the vast majority of cases sensitive ad preferences are assigned based on the activity of the user (e.g., page like, ad click) rather than an act of the user to ``manifestly made public" some sensitive personal data. We have listed in the paper the reasons FB uses to explain why an ad preference has been assigned. None of these reasons is associated with the public disclosure of personal data in the user profile or wall. Hence, this exception does not apply to the vast majority of users in FB.

(f) \textit{``processing is necessary for the establishment, exercise or defense of legal claims or whenever courts are acting in their judicial capacity."}

This exception does not apply to FB since it is very unlikely that ad preferences can be used to the establishment, exercise or defense or legal claims. Ad preferences are used for advertising purposes.

(g) \textit{``processing is necessary for reasons of substantial public interest, on the basis of Union or Member State law which shall be proportionate to the aim pursued, respect the essence of the right to data protection and provide for suitable and specific measures to safeguard the fundamental rights and the interests of the data subject."}

FB processes sensitive personal data for commercial purposes. The processing of sensitive personal data in the case of FB has nothing to do with public interest. Hence, this exception does not apply to the case of FB.

(h) \textit{``processing is necessary for the purposes of preventive or occupational medicine, for the assessment of the working capacity of the employee, medical diagnosis, the provision of health or social care or treatment or the management of health or social care systems and services on the basis of Union or Member State law or pursuant to contract with a health professional and subject to the conditions and safeguards referred to in paragraph 3\footnote{Paragraph 3 can be found in \cite{GDPR}}."}

FB operation has nothing to do with any of the premises included in this exception. Therefore, this exception does not apply to the case of FB.

(i) \textit{``processing is necessary for reasons of public interest in the area of public health, such as protecting against serious cross-border threats to health or ensuring high standards of quality and safety of healthcare and of medicinal products or medical devices, on the basis of Union or Member State law which provides for suitable and specific measures to safeguard the rights and freedoms of the data subject, in particular professional secrecy."}

FB operation, and in particular the exploitation of ad preferences, is not related to the area of public health. Therefore, this exception does not apply to the case of Facebook.

(j) \textit{``processing is necessary for archiving purposes in the public interest, scientific or historical research purposes or statistical purposes in accordance with Article 89(1) based on Union or Member State law which shall be proportionate to the aim pursued, respect the essence of the right to data protection and provide for suitable and specific measures to safeguard the fundamental rights and the interests of the data subject"}

FB sensitive ad preferences are (and should) not be publicly shared. In addition, processing sensitive data from FB users is neither critical for scientific nor for statistical purposes. Therefore, this exception does not apply to the case of Facebook.

Therefore, according to the GDPR text covering sensitive personal data processing, the practice of labeling FB users with ad preferences associated with sensitive personal data may be contravening the article 9 of the GDPR.

\subsection{Facebook fined in Spain}
\label{subsec:aepd_fine}

In September 2017 the Spanish Data Protection Agency (AEPD) fined Facebook \euro1.2M for violating the data protection regulation. In the fine's resolution \cite{AEPD} the AEPD claims that FB collects, stores and processes sensitive personal data for advertising purposes without obtaining consent from the users. 

In the following, we literally list the main elements included in the Spanish DPA resolution associated with the \euro1.2M fine imposed to FB for violating the Spanish data protection regulation.
\begin{itemize}
\item[-] \textit{The Agency notes that the social network collects, stores and uses data, including specially protected data, for advertising purposes without obtaining consent.}

\item[-] \textit{The data on ideology, sex, religious beliefs, personal preferences or browsing activity are collected directly, through interaction with their services or from third party pages without clearly informing the user about how and for what purpose will use those data.}

\item[-] \textit{Facebook does not obtain unambiguous, specific and informed consent from users to process their data since the information it offers is not adequate}

\item[-] \textit{Users' personal data are not totally canceled when they are no longer useful for the purpose for which they were collected, nor when the user explicitly requests their removal.}

\item[-] \textit{The Agency declares the existence of two serious and one very serious infringements of the Data Protection Law and imposes on Facebook a total sanction of 1,200,000 euros.}

\item[-] \textit{The AEPD is part of a Contact Group together with the Authorities of Belgium, France, Hamburg (Germany) and the Netherlands, that also initiated their respective investigation procedures to the company.} 
\end{itemize}

This fine also confirms that currently enforced data protection regulations in some EU countries (that will be substituted by the GDPR in May 2018) are already considering inappropriate the processing and storage of sensitive personal data without an unambiguous and specific informed consent of the user. Therefore, FB may have been contravening data protection laws of several EU countries for some time.

It is worth noting that also in US Facebook was involved in a civil rights lawsuit for allowing advertisers to target people based on the attribute \textit{ethnic affinity}. FB renamed such attribute to \textit{multicultural affinity} and accepted that it could not be combined in campaigns targeting housing, employment and financial services \cite{angwin_2016}\cite{fbresp}\cite{angwin_2017}\cite{sandberg_2017}. 


\subsection{Facebook Terms of Service}
\label{subsec:fb_ts}
We have carefully revised the terms of service and data policy subscribed by users and advertisers on Facebook. 

EU FB users consent FB to transfer to and process personal data in the United States. However, the GDPR (and other regulations in Europe) clearly differentiate between personal data and some categories of the so-called \textit{specially protected} or \textit{sensitive} personal data. In addition, the referred regulations require obtaining explicit permission from the user to process and store sensitive personal data. Still, after our thorough review of FB terms of service and data policy, we could neither find a clear piece of text where EU users are informed that FB is processing and storing \textit{sensitive personal data} according to the GDPR definition, nor a place where users provide explicit consent for the processing of this data. 

FB defines in its advertising policy that ads must not include content that asserts or implies personal attributes. By personal attributes, FB refers to most of the categories of sensitive data listed in the GDPR. Facebook provides examples of what content in an ad is considered appropriate and what is not. For instance, \textit{Are you gay?}, \textit{Do you have diabetes?} and \textit{Meet other Buddhists} are not accepted, whereas \textit{Gay dating online now}, \textit{New diabetes treatment available} and \textit{Looking for Buddhists near you?} are valid. Therefore, Facebook prohibits to use content that directly implies to acknowledge some particular sensitive personal information from the targeted user, but not indirect implications associated with the content. Surprisingly, the ad received by one of the authors referred in the Introduction (see Figure \ref{fig:ad1}) was actually contravening FB rules. The text included in the ad was: \textit{Connect with the gay community \& rent affordable places from people like you. Book Now}. By using the expression \textit{``like you"}, the advertising is explicitly indicating that the author of the paper is gay. Finally, it is important to note that FB ads policy does not prevent by any means that advertisers target FB users based on sensitive personal data, but only regulates the content of the delivered ad.

FB users subscribe Facebook Terms of Service\footnote{\url{https://www.facebook.com/terms.php}} when opening a FB account. This is the entry document where users are informed what FB is doing with their personal data. 
However, in order to better understand the details regarding FB data management users are redirected to another document referred to as Data Policy\footnote{\url{https://www.facebook.com/about/privacy/}}. We  found three sections very relevant for our research in the Terms of Service document: \\

\noindent \textbf{Section 16. Special Provisions Applicable to Users Outside the United States.} This section includes the following clause \textit{``You consent to have your personal data transferred to and processed in the United States."} While these grants DB sufficient permission to process and store personal data, the GDPR and some precedent data protection regulations in some EU countries establish a clear difference between personal data and \textit{``specially protected"} or \textit{``sensitive"} personal data. To the best of our knowledge, FB does not obtain explicit permission to process and store sensitive personal data.\\

\noindent \textbf{Section 9. About Advertisements and Other Commercial Content Served or Enhanced by Facebook.} In this section, users are informed that FB can use the user information, name, picture, etc. for advertising and commercial purposes. Therefore, it is true that FB states that users' information will be used for commercial purposes.\\

\noindent \textbf{Section 10. Special Provisions Applicable to Advertisers }. Advertisers are forwarded to two more documents Self-Serve Ad Terms\footnote{\url{https://www.facebook.com/legal/self\_service\_ads\_terms}} (not very relevant for our research) and Advertising Policies\footnote{\url{https://www.facebook.com/policies/ads/}}. The latter document includes 13 sections from which Section 4.12\footnote{\url{https://www.facebook.com/policies/ads/prohibited\_content/personal\_attributes}} (4-Prohibited Content; 12- Personal attributes) is very relevant for our paper. Section 4.12 literally states: \textit{``Ads must not contain content that asserts or implies personal attributes. This includes direct or indirect assertions or implications about a person’s race, ethnic origin, religion, beliefs, age, sexual orientation or practices, gender identity, disability, medical condition (including physical or mental health), financial status, membership in a trade union, criminal record, or name."}. Examples of what content is allowed and what content is prohibited are provided in the Advertising Policies.
\section{Dataset}
In order to analyze the presence of sensitive ad preferences and quantify the portion of EU FB users that have been assigned some of them, we should be able to collect a large dataset of ad preferences assigned to users. Hence, in case we detect some ad preferences that represent \textit{sensitive personal data} we have the proof that they have been assigned to real FB users. Based on this principle our dataset is created from the ad preferences collected from real users who have installed the FDVT. We note that the number of ad preferences retrieved from the FDVT represents just a subset of the overall set of preferences, but we can guarantee that they have been assigned to real users.
Our dataset includes the ad preferences from 4577 users who installed the FDVT between October 2016 and October 2017, from which 3166 users come from some EU country. The 4577 FDVT users have been assigned in total 5.5M ad preferences from which we derive 126192 unique. 


Our dataset includes the following information for each ad preference:
\begin{itemize}
 
\item[-]\textit{\underline{ID of the ad preference}}: This is the key we use to identify an ad preference independently of the language used by a FB user. For instance, the ad preference \{Milk, Leche, Lait\} that refers to the same thing in English, Spanish and French, is assigned a single FB ID. Therefore, we can uniquely identify each ad preference across all EU countries and languages.

\item[-]\textit{\underline{Name of the ad preference}}: This is the primary descriptor of the ad preference. FB returns a unified version of the name for each ad preference ID, usually in English. Hence, we have the English name of the ad preferences irrespectively of the original language at the collection time. We note that in some cases translating the ad preference name does not make sense (e.g., the case of persons' names: celebrities, politicians, etc.). 

\item[-]\textit{\underline{Disambiguation Category}}: For some ad preferences Facebook adds this in a separate field or in parenthesis to clarify the meaning of a particular ad preference 
(e.g., Violet (color); Violet: Clothing (Brand))
We have identified more than 700 different disambiguation category topics (e.g., Political Ideology, Disease, Book, Website, Sport Team, etc.).

\item[-]\textit{\underline{Topic Category}}: Some of the 14 first level interests introduced in Section \nameref{subsec:ads_manager} are assigned in many cases to contextualize ad preferences. For instance, Manchester United F.C. is linked to Sports and Outdoors.

\item[-]\textit{\underline{Audience Size}}: This value reports the number of Facebook users that have been assigned with the ad preference worldwide.

\item[-]\textit{\underline{Reason why the ad preference is added to the user}}: The reason why the ad preference has been assigned to the user according to FB. There are 6 possible reasons already introduced in Subsection \nameref{subsec:ads_preferences}.
\end{itemize}
\begin{figure*}[t]
\centering
	\begin{minipage}[t]{0.32\hsize}
		\centering
		\includegraphics[width=1\columnwidth]{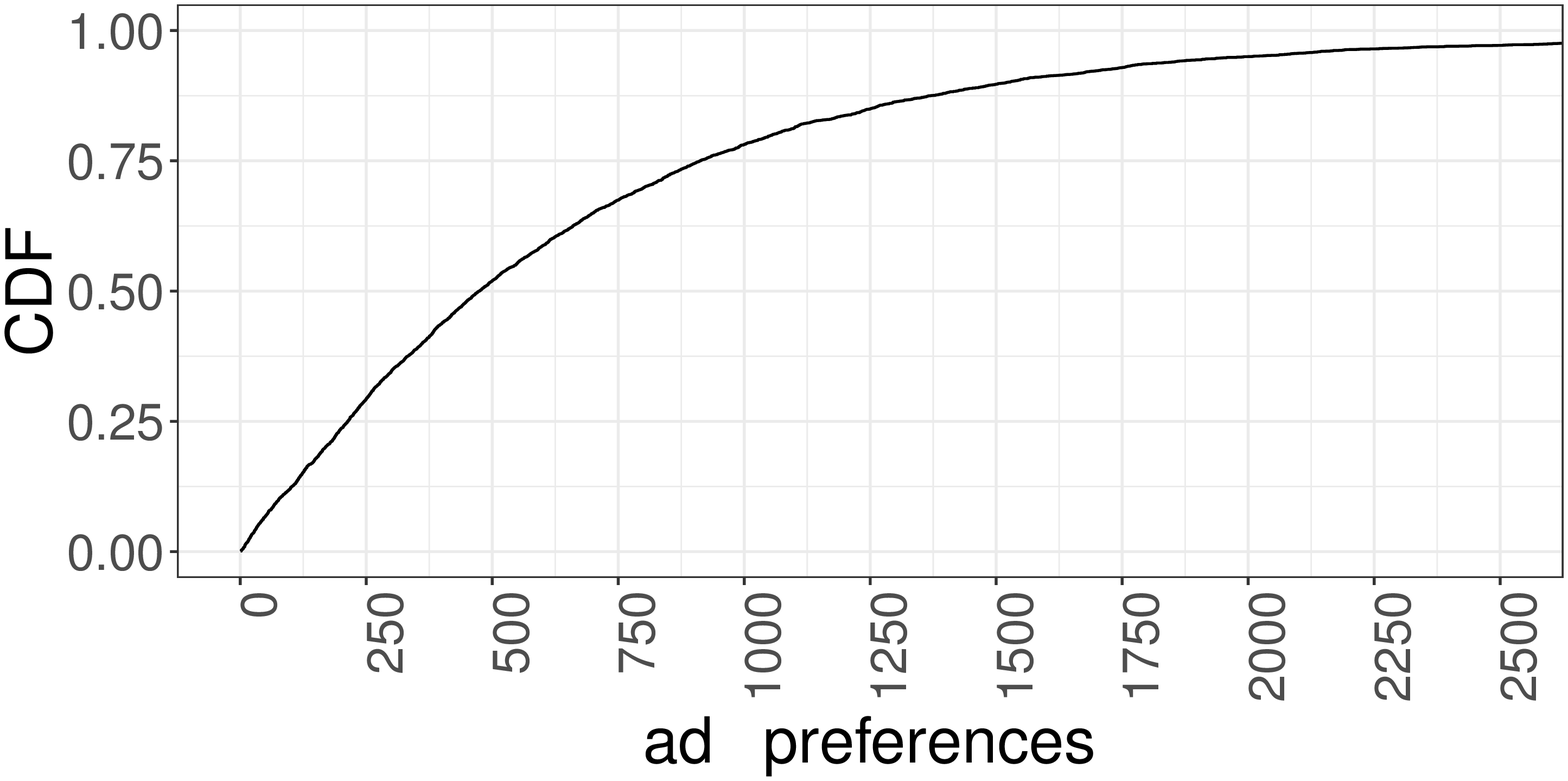}\hfill
		\caption{CDF of the number of ad preferences per FDVT user.}
		\label{fig:ecdf}
	\end{minipage}
    \hfill
	\begin{minipage}[t]{0.32\hsize}
		\centering
		\includegraphics[width=1\columnwidth]{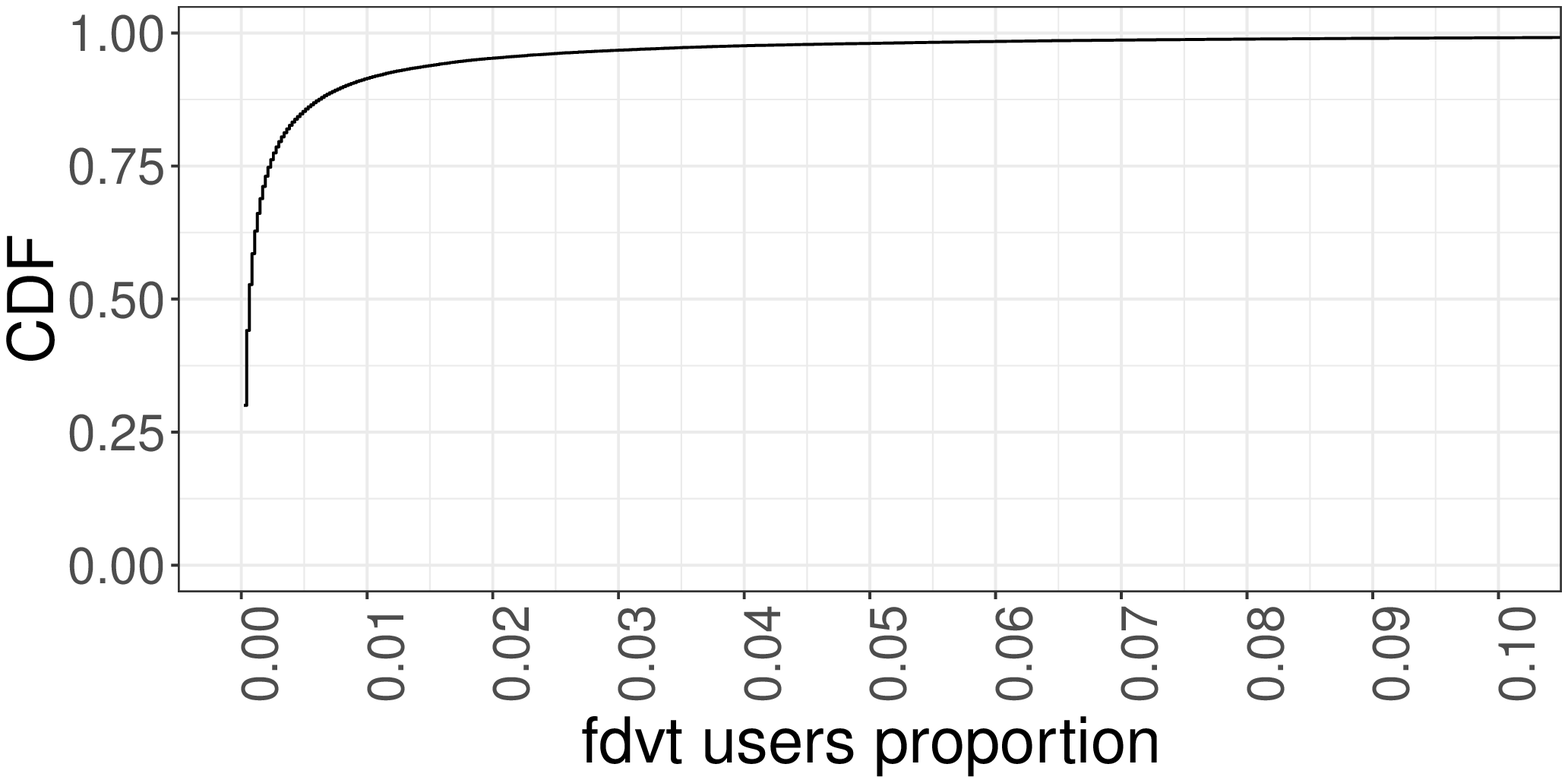}\hfill
		\caption{CDF of the portion of FDVT users (x-axis) per ad preference (y-axis).}
		\label{fig:usersXpref}
    \end{minipage}
    \hfill
    \begin{minipage}[t]{0.32\hsize}
		\centering
		\includegraphics[width=1\columnwidth]{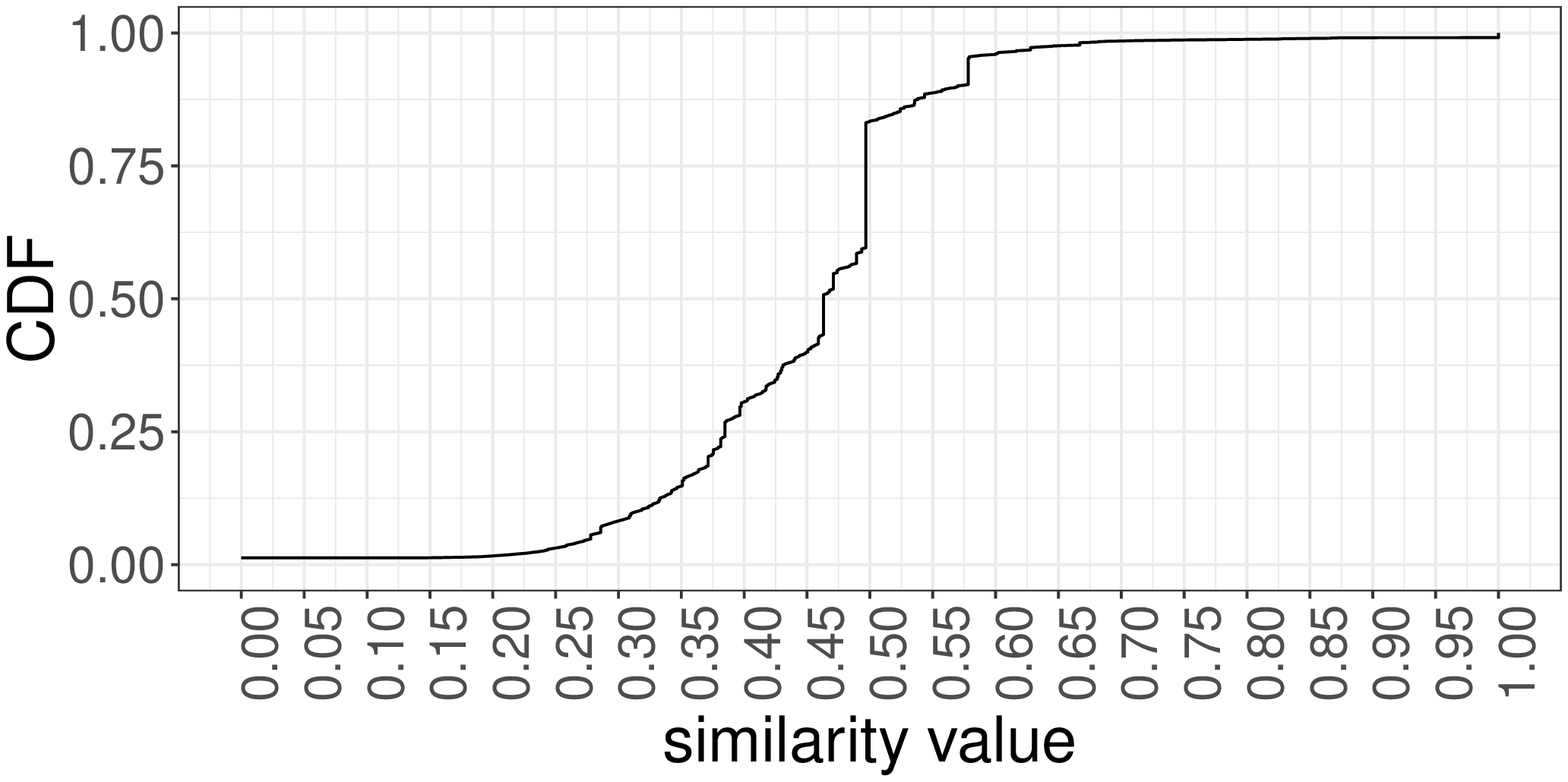}\hfill
		\caption{CDF of the semantic similarity score assigned to the 126K ad preferences from the FDVT dataset.}
		\label{fig:cdf_clasificacion}
	\end{minipage}
\end{figure*}

Figure \ref{fig:ecdf} shows the CDF of the number of ad preferences per user. In median, each FDVT user is assigned 474 preferences. Moreover, Figure \ref{fig:usersXpref} shows the CDF of the portion of FDVT users (x-axis) that got assigned a given ad preference (y-axis). We observe a very skewed distribution that indicates that most ad preferences are actually assigned to a small fraction of users. For instance, in median, an ad preference is assigned to only 3 (0.06\%) FDVT users. However, it is important to note that still many ad preferences reach a relevant portion of users. Our dataset includes 1000 ad preferences that reach at least 11\% FDVT users. 


\section{Methodology}
\label{sec:methodology}
Our final goal is to quantify the number of EU FB users that have been assigned sensitive ad preferences. To this end, we use as reference the 126K unique ad preferences assigned to FDVT users and follow a two-step process. In the first step, we combine Natural Language Processing (NLP) techniques with human manual classification to obtain the list of sensitive ad preferences from the 126K considered. In the second step, we leverage the FB Ads Manager API to quantify how many FB users in each EU country have been assigned at least one of the ad preferences labeled as sensitive.  

\subsection{Identification of Sensitive Ad Preferences}

We rely on a group of researchers with some knowledge in the area of privacy to manually identify the sensitive ad preferences within our pool of 126K ad preferences retrieved from FDVT users. However, classifying manually  126K ad preferences does not scale\footnote{If we consider 10s as the average time required to classify an ad preference as sensitive vs. non-sensitive, a person would need roughly 44 full working days of 8 hours only dedicated to this task.}. To make this manual classification task scalable, we leverage NLP techniques to pre-filter the list of ad preferences more likely to be sensitive. This pre-filtering phase will deliver a subset of likely sensitive ad preferences that can be manually classified in a reasonable amount of time.

\subsubsection{Pre-filtering}

\paragraph{\textbf{Sensitive Categories}}

To identify likely sensitive ad preferences in an automatic manner, we select five of the most relevant categories listed as \textit{Sensitive Personal Data} by the GDPR: $(i)$ data revealing racial or ethnic origin, $(ii)$ data revealing political opinions, $(iii)$ data revealing religious or philosophical beliefs, $(iv)$ data concerning health, and $(v)$ data concerning sex life and sexual orientation. We selected these categories because a preliminary manual inspection indicated that there are ad preferences in our dataset that can likely reveal information related to them. For instance, the ad preferences \textit{``Socialism"},\textit{``Islam"},\textit{``Reproductive Health"},\textit{``Homosexuality"},\textit{``Chinese culture"} may reveal \textit{political opinion, religious belief, health issue, sexual orientation, ethnic or racial origin} of the users that have been assigned them, respectively. Note that all these examples of ad preferences have been extracted from our dataset; thus they have been assigned to actual FB users.

Our automatic process will classify an ad preference as \emph{likely sensitive} if we can semantically map that ad preference name into one of the five sensitive categories analyzed in this paper. To this end, we have defined a dictionary including both keywords and short sentences representative of each of the five considered sensitive categories. We have used two data sources to create the dictionary: First, a list of controversial issues available in Wikipedia\footnote{\url{https://en.wikipedia.org/wiki/Wikipedia:List_of_controversial_issues}}. In particular, we have selected the following categories from this list: politics and economics, religion and, sexuality. Second, we have obtained a list of words with a very similar semantic meaning to the five sensitive personal data categories. To this end, we have used the Datamuse API\footnote{\url{https://www.datamuse.com/api/}}, a word-finding query engine that allows developers to find words that match a set of constraints. Among other functionalities, Datamuse allows \textit{``finding words with a similar meaning to X"} using a simple query. 

The final dictionary includes 264 keywords\footnote{\url{https://fdvt.org/usenix2018/keywords.html}}. We leverage the keywords in this dictionary to find ad preferences that present high semantic similarity to at least one of these keywords. In such case, we tag them as likely sensitive ad preferences. It is worth noting that this approach makes our methodology flexible since the dictionary can be extended to include new keywords for the already considered categories, or related to other sensitive categories.

We next describe the semantic similarity computation in detail.

\paragraph{\textbf{Semantic Similarity Computation}}

The semantic similarity computation process takes two inputs: the 126K ad preferences from our FDVT dataset and the 264 keywords dictionary associated with the considered sensitive categories. We compute the semantic similarity of each ad preference with all the 264 keywords from the dictionary. For each ad preference, we record the highest similarity value out of the 264 comparison operations. As result of this process, each one of the 126K ad preferences is assigned a similarity score, which indicates its likelihood to be a sensitive ad preference.

To implement the semantic similarity comparison task, we leverage the Spacy package for python\footnote{\url{https://spacy.io}} 

Spacy is a free open source package for advance NLP operations. Spacy offers multiple NLP features such as information extraction, natural language understanding, deep learning for text, semantic similarity analysis, etc., which are accomplished through different predefined models. To conduct our analysis, we leverage the ``similarity" feature of Spacy that allows comparing two words or short text providing a semantic similarity value ranging between 0 (lowest) and 1 (highest). This feature computes similarity using the so-called Glove (Global vectors for word representation)  method \cite{pennington2014glove}. Gloves are multi-dimensional meaning representations of words computed using word2vec \cite{mikolov2013efficient}\cite{mikolov2013distributed}\cite{mikolov2013linguistic}. Spacy word vectors are trained using a large corpus of text incorporating a rich vocabulary. In addition, Spacy also takes into account context to define the representation of a word, which allows to better identify its meaning considering the surrounding words. Spacy offers different models to optimize the semantic similarity computation. We have chosen the model \textit{en\_core\_web\_md}\footnote{\url{https://spacy.io/models/en\#en\_core\_web\_md}} because it optimizes the similarity analysis between words and short sentences, which matches the nature of ad preferences names. The chosen model is an English multi-task Convolutional Neural Network (CNN) trained on OntoNotes \cite{weischedel2013ontonotes} with GloVe vectors that are in turn trained on Common Crawl \footnote{\url{http://commoncrawl.org/}}. Common Crawl is an open source repository for crawling data. The model uses word vectors, context-specific token vectors, POS (part-of-speech) tags, dependency parse and named entities.

We have chosen Spacy because it has been previously used in the literature for text processing purposes offering good performance \cite{korpusik2017semantic}\cite{panchenko2016best}. Moreover, Spacy offers good scalability. It computes the 33314688 (126192 x 264) semantic similarity computations in 7 min using a server with twelve 2.6GHZ cores and 96GB of RAM. To conduct our analysis we leverage the \textit{similarity} feature of Spacy. This feature allows comparing words,  text spans or documents, and computes the semantic similarity among them. The output is a semantic similarity value ranging between 0 and 1. The closer to 1 the higher the semantic similarity is.

The execution of the explained process revealed very low similarity values in some cases in which the analyzed ad preference matched pretty well the definition of some of the sensitive personal data categories. Some of these cases are: physical persons such as politicians (which may reveal the political opinion of the user); political parties with names that do not include any standard political term; health diseases or places of religious cult that may have names with low semantic similarity with health and religious related keywords in our dictionary, respectively. Three examples illustrating the referred cases are: $<$name: ``Angela Merkel", disambiguation: Politician$>$; $<$name: ``I Love Italy", disambiguation: Political Party$>$; $<$name: ``Kegel" exercise, disambiguation: Medical procedure$>$. 
We realized that in most of these cases the disambiguation category is actually a better element than the ad preference name to perform the semantic similarity analysis. For instance, in the case of politicians names, political parties and health diseases the disambiguation category field includes the term \textit{``politician"}, \textit{``Political Party"} and \textit{``disease"}, respectively. The disambiguation category field is also very useful for ad preference names that have multiple meanings since it allows to determine which of those meanings a specific ad preference is actually referring to. 

Overall, we found that for classifying ad preferences, the disambiguation category, when it is available, is a better proxy than the ad preference name. Therefore, in case the ad preference under analysis has associated a disambiguation category field, we used the disambiguation category string instead of the ad preference name to obtain the semantic similarity score of the ad preference.
\\

\paragraph{\textbf{Selection of likely sensitive ad preferences}}

The semantic similarity computation process assigns a similarity score to each one of the 126K ad preferences in our dataset. This similarity score represents the likelihood for an ad preference to be sensitive.

In this step of the process, we have to select a relatively high similarity score threshold that allows us to create a subset of likely sensitive ad preferences that can be manually labeled using a reasonable amount of resources (i.e., number of persons, and time required per person to label ad preferences).

Figure \ref{fig:cdf_clasificacion} shows the CDF for the semantic similarity score of the 126K ad preferences. The curve is flat in its initial and final part, with a big rise between similarity values 0.25 and 0.6. This big rise implies that setting our threshold in values below 0.6 would result in a rapid growth of the number of ad preferences to be manually tagged. Therefore, we have decided to set up the semantic similarity threshold in 0.6 because it corresponds to a relatively high similarity score, and the resulting subset includes 4452 ad preferences (3.5\% out of the 126K), which is an affordable number to be manually tagged.

Note that the CDF shows two jumps at similarity scores equal to 0.5 and 0.58. The first one is linked to the disambiguation category \textit{``Local Business"} while the second one refers to the disambiguation category \textit{``Public Figure"}. Overall, we do not expect to find a significant number of sensitive ad preferences within these disambiguation categories. Hence, this observation reinforces our semantic similarity threshold selection of 0.6.


\subsubsection{Manual Classification of Sensitive Ad Preferences}
\label{subsec:manual_classification}

We have recruited twelve panelists. All of them are researchers (faculty and Ph.D. students) with some knowledge in the area of privacy. Each panelist has manually classified a random sample (between 1000 and 4452 elements) from the 4452 ad preferences included in the filtered dataset described above. We asked them to classify each ad preference into one of the five considered sensitive categories (Politics, Health, Ethnicity, Religion, Sexuality), in the category ``Other" (if it does not correspond to any of the sensitive categories), or in the category ``Not known" (if the panelist does not know the meaning of the ad preference). To carry out the manual labeling of ad preferences the researchers were given all the contextual information Facebook offers per ad preference: ad preference name, disambiguation category (if it is available) and topic (if it is available). The literal instructions given to the researchers to complete the classification task were: \textit{``Assign only one category per ad preference. If you think that more than one category applies to an ad preference use only the one you think is more relevant. If none of the categories match the ad preference, classify it as `Other'. In case you do not know the meaning of an ad preference please read the disambiguation category and topic that may help you. If after reading them you still are unable to classify the ad preference, use `Not known' to classify it."} 

Each ad preference has been manually classified by five panelists. We use majority voting \cite{narasimhamurthy2005theoretical} to classify each ad preference either as sensitive or non-sensitive. This is, we label an ad preference as sensitive if at least 3 voters (i.e., the majority) classify it in one of the 5 sensitive categories and as non-sensitive otherwise.

\begin{table}[t]
\centering
\begin{adjustbox}{width=0.40\textwidth}
\begin{tabular}{rrrrrrr}
  \hline votes & 0 & 1 & 2 & 3 & 4 & 5 \\ 
  \hline
  \#preferences & 1054 & 767 & 539 & 422 & 449 & 1221  \\ 
   \hline
\end{tabular}
\end{adjustbox}
\caption{Number of ad preferences that have received 0, 1, 2, 3, 4 or 5 votes classifying them into one sensitive data categories. }
\label{table:votes}
\end{table}

Table \ref{table:votes} shows the number of ad preferences that have received 0, 1, 2, 3, 4 and 5 votes classifying them into a sensitive category. 2092 out of the 4452 ad preferences are labeled as sensitive, i.e., have been classified into a sensitive category by at least 3 voters. This represents 1,66\% of the 126K ad preferences from our dataset.

An ad preference classified as sensitive may have been assigned to different sensitive categories (e.g., politics and religion) by different voters. We have evaluated the voters' agreement across the sensitive categories assigned to ad preferences labeled as sensitive using the Fleiss' Kappa test \cite{fleiss1971measuring}\cite{fleiss2013statistical}.  The Fleiss' Kappa coefficient obtained is 0.94. This indicates an almost perfect agreement among the panelists' votes that link an ad preference to a sensitive category\cite{landis1977measurement}. Hence, we conclude that (almost) every ad preference classified as sensitive corresponds to a unique sensitive category among the 5 considered.

The 2092 ad preferences manually labeled as sensitive are distributed as follows across the five sensitive categories: 58.3\% are related to politics, 20.8\% to religion, 18.2\% to health, 1.5\% to sexuality,  1.1\% to ethnicity and just 0.2\% present discrepancy among votes. The complete list of the ad preferences classified as sensitive can be accessed in\footnote{\url{https://fdvt.org/usenix2018/panelists.html}}.



\subsection{Retrieving the number of FB users assigned sensitive ad preferences from the FB Ads Manager}

We leverage the FB Ads Manager API to retrieve the number of FB users in each EU country that have been assigned each of the 2092 ad preferences labeled as sensitive. Besides, we have sorted the sensitive ad preferences from the most to the least popular in each country. This allows us to compute the number of FB users assigned at least with one of the Top N sensitive ad preferences (with N ranging between 1 and 2092). To obtain this information we use the OR operation available in the FB Ads Manager API to create audiences. This feature allows us to retrieve how many users in a given country are interested in \textit{ad preference 1} OR \textit{ad preference 2} OR \textit{ad preference 3}... OR \textit{ad preference N}.  An example of this for N = 3 could be  \textit{``how many people in France are interested in Communism OR Islam OR Veganism"}. 

Although the number of users is a relevant metric, it does not offer a fair comparative result to assess the importance of the problem across countries because we can find EU countries with tens of millions of users (e.g., France, Germany, Italy, etc) and some others with less than a million (e.g., Malta, Luxembourg, etc). Hence, we will use the portion of users in each country that have been assigned sensitive ad preferences as the metric to analyze the results. Beyond FB users we are also interested in quantifying the portion of citizens assigned sensitive ad preferences in each EU country. We have defined two metrics used in the rest of the paper:  
\begin{itemize}
\item [-]\emph{\textbf{FFB(C,N)}}: This is the percentage of FB users in country C that have been assigned at least one of the top N sensitive ad preferences. We note C may also refer to all 28 EU countries together when we want to analyze the results for the whole EU. It is computed as the ratio between the number of FB users that have been assigned at least one of the top N sensitive ad preferences and the total number of FB users in country C, which can be retrieved from the FB Ads Manager.

\item [-]\emph{\textbf{FC(C,N)}}: This is the percentage of citizens in country C (or all EU countries together) that have been assigned at least one of the top N sensitive ad preferences. It is computed as the ratio between the number of citizens that have been assigned at least one of the top N sensitive ad preference and the total population of country C. We use World Bank data to obtain EU countries population\footnote{\url{https://data.worldbank.org/indicator/SP.POP.TOTL?locations=EU}}.
\end{itemize}
Note that FFB(C,N) and FC(C,N) will report a lower bound concerning the total percentage of FB users and citizens in country C tagged with sensitive ad preferences for two reasons. First, these metrics can use at most N = 2092 sensitive ad preferences, which is very likely a subset of all sensitive ad preferences available on FB. Second, the FB Ads Manager API only allows creating audiences with at most 1000 interests (i.e., ad preferences). Beyond 1000 interests the API provides a fixed number of FB users independently of the defined audience. This fixed number is 2.1B which to the best of our knowledge refers the total number of FB users included in the Ads Manager. Therefore, in practice, the maximum value of N we can use in FFB and FC is 1000.

\section{Quantifying the exposition of EU users to sensitive ad preferences}
\label{sec:results}
\begin{table}[t]
\centering
\begin{adjustbox}{width=1\columnwidth}
\begin{tabular}{rcc}
  \hline
 reason of assignment & all ad preferences & sensitive ad preferences \\ 
  \hline
  due to a like & 71.64\% & 81.36\% \\
  due to an ad click & 21.51\% & 15.85\% \\
  FB suggests it could be relevant & 4.83\% & 2.45\% \\
  due to an app installation & 1.78\% & 0.04\% \\
  due to comments or reaction buttons & 0.18\% & 0.26\% \\
  added by user & 0.04\% & 0.03\% \\ 
  unclear or not gathered by FDVT & 0.01\% & 0.01\% \\  
   \hline
\end{tabular}
\end{adjustbox}
\caption{Frequency of the 6 reasons why ad preferences are assigned to FDVT EU users according to FB explanations.}
\label{table:reasons}
\end{table}

In this section, we first analyze the exposition of the FDVT users to the 2092 sensitive ad preferences found in the previous section. Afterwards, we use the FFB and FC metrics to analyze the exposition of EU FB users and citizens to those ad preferences. Finally, we perform a demographic analysis to understand whether users from certain gender or age groups are more exposed to sensitive ad preferences.
\vspace{2em}

\subsection{FDVT Users}

4121 (90\%) FDVT users are tagged with at least one sensitive ad preference. Overall, the 2092 unique sensitive ad preferences have been assigned more than 146K times to the FDVT users. If we focus only on EU users, which are the subject of study in this paper, 2848 (90\%) have been tagged with sensitive ad preferences. Overall, they have been assigned more than 100K sensitive interests (1528 unique). The median (avg) number of sensitive ad preferences assigned to FDVT users is 10 (16). The 25th and 75th percentiles are 5 and 21, respectively.

Our FDVT dataset includes the reason why, according to FB, each ad preference has been assigned to a user. Table \ref{table:reasons} shows the frequency of each reason for both all ad preferences and only sensitive ad preferences. The results indicate that most of the sensitive ad preferences are derived from \textit{users likes} (81\%) or \textit{clicks on ads} (16\%). There are very few cases (0.03\%) in which users proactively include sensitive ad preferences in their list of ad preferences using the configuration setting offered by FB. We remind that, according to the EU GDPR, Facebook should obtain explicit permission to process and exploit sensitive personal data. Users likes and clicks on ads do not seem to meet this requirement. 

\subsection{EU FB users and citizens}

\begin{figure*}[t]
\centering
  \begin{minipage}[t]{0.32\hsize}
      \centering
      \includegraphics[width=1\textwidth]{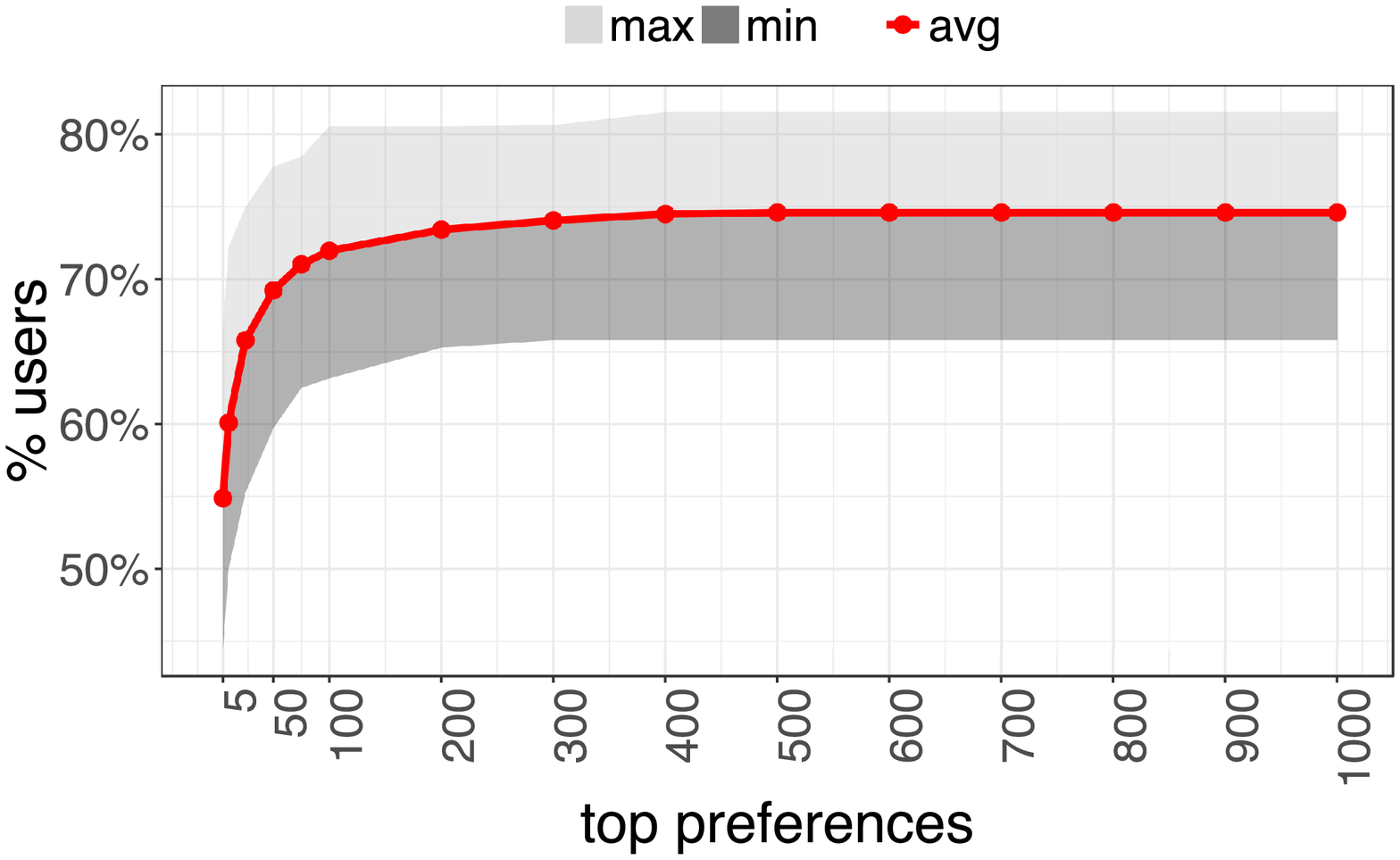}\hfill
      \caption{FFB (C,N) for values of N ranging between 1 and 1000. The figure reports the min, average and max FFB value across the 28 EU countries.}
      \label{fig:tops_FB}
  \end{minipage}
  \hfill
  \begin{minipage}[t]{0.32\hsize}
      \centering
      \includegraphics[width=1\textwidth]{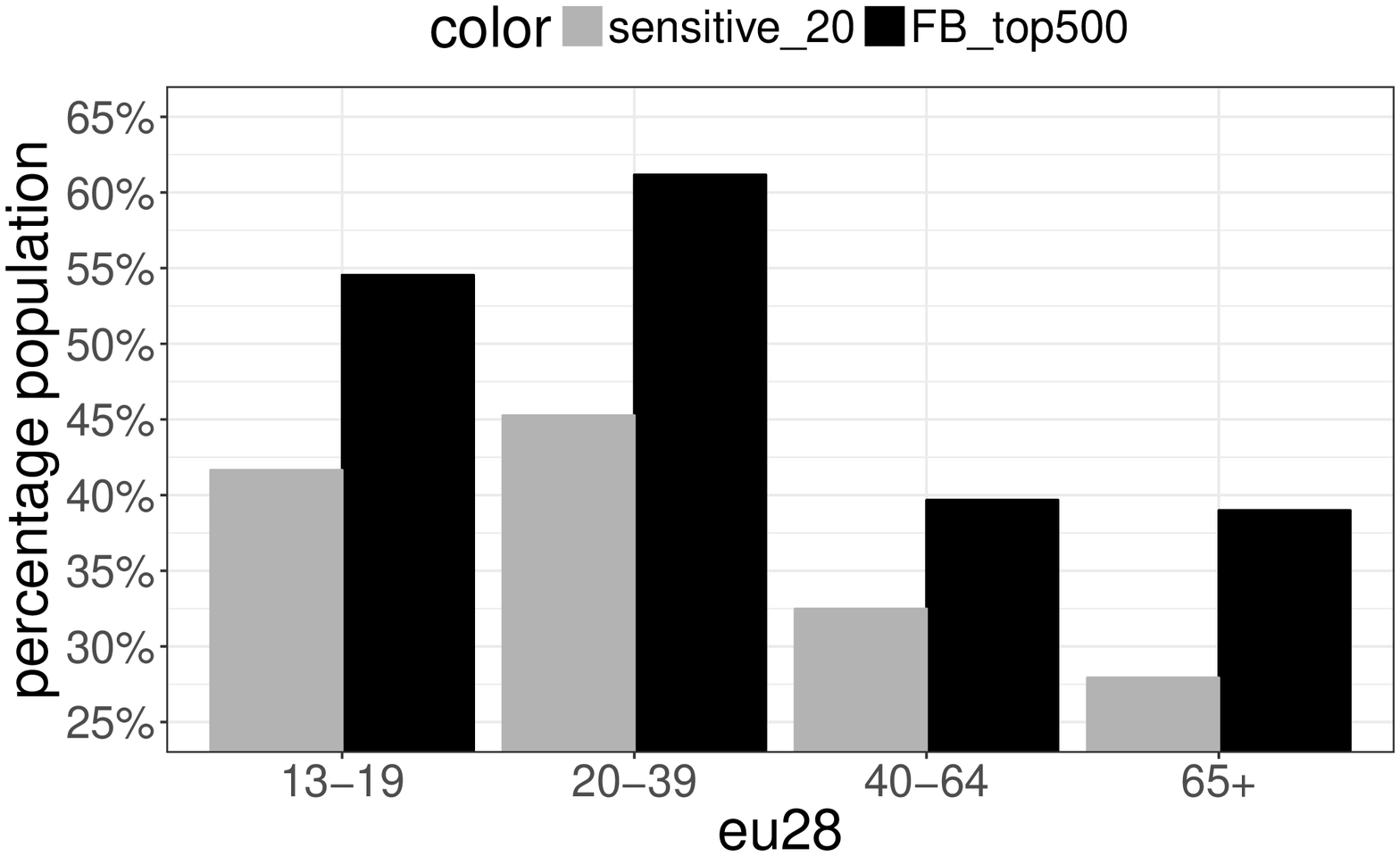}\hfill
      \caption{Percentage of EU FB users assigned at least one of the Top 500 (black) and 20-very sensitive (grey) ad preferences in the following age groups: 13-19, 20-39, 40-64, 65+.}
      \label{fig:age_groups}
  \end{minipage}
  \hfill
  \begin{minipage}[t]{0.32\hsize}
      \centering
      \includegraphics[width=1\textwidth]{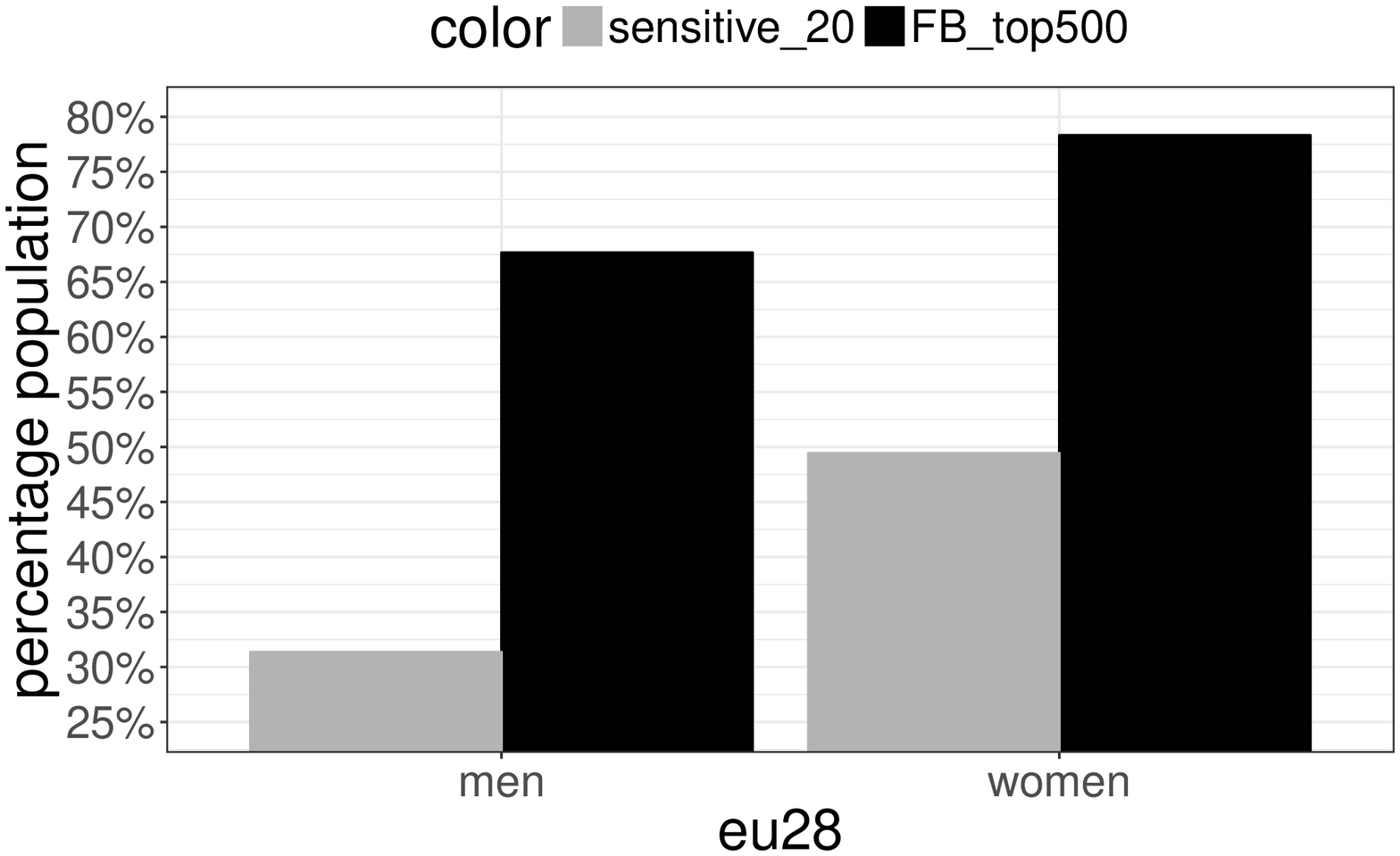}\hfill
      \caption{Percentage of EU FB users assigned at least one of the Top 500 (black) and 20-very sensitive (grey) ad preferences in the following gender groups: Men, Women.}
      \label{fig:genders_groups}
  \end{minipage}
\end{figure*}

Figure \ref{fig:tops_FB} shows the FFB (C,N) for values of N ranging between 1 and 1000. The figure reports the max, min and avg values across the 28 EU countries. We observe that even if we consider a low number of sensitive ad preferences, the fraction of affected users is very significant. For instance, in average 60\% FB users from EU countries are tagged with some of the top 10 (i.e., most popular) sensitive ad preferences.

Moreover, we observe that FFB is stable for values of N ranging between 500 and 1000. We note that we have obtained the same stable result for each individual EU country. This indicates
that any user tagged with sensitive ad preferences outside the top 500\footnote{The top 500 list by country can be accessed at \url{https://fdvt.org/usenix2018/top500.html}} has been very likely already tagged with at least one sensitive ad preference within the top 500. We conjecture that this asymptotic behavior may indicate that the lower bound represented by FFB(C, N=500) is close to the actual fraction of FB users tagged with sensitive ad preferences. 


Table \ref{table:top_interests} shows FFB(C,N=500) and FC(C,N=500) for every EU country. The last row in the table shows aggregated results for the 28 EU countries together (EU28). 

We observe that 73\% EU FB users, that corresponds to 40\% EU citizens, are tagged with some of the top 500 sensitive ad preferences in our dataset. If we focus on individual countries, FC(C,N=500) reveals that in 7 of them more than half of their citizens are tagged with at least one of the top 500 sensitive ad preferences: Malta (66.37\%), Cyprus (64.95\% ), Sweden (54.53\%), Denmark (54.09\%), Ireland (52.38\%), Portugal (51.33\%) and Great Britain (50.28\%). In contrast, the 5 countries least impacted are: Germany (30.24\%), Poland (31.62\%), Latvia (33.67\%), Slovakia (35\%) and Czech Republic (35.98\%). Moreover, FFB(C,N=500) ranges between 65\% for France and 81\% for Portugal. This means that at least 2/3 FB users in any EU country are tagged with some of the top 500 sensitive ad preferences.

These results draw a worrisome situation where a very significant part of the EU population is exposed to being targeted by advertising campaigns based on sensitive personal data. 


\begin{table}[t]
\centering
\begin{adjustbox}{width=1\columnwidth}
\begin{tabular}{llrr|llrr}
  \hline
country & C & FFB(C,500) & FC (C,500) & country & C & FFB(C,500) & FC (C,500)\\ 
  \hline
Austria & AT & 75.00 & 37.73  &  Ireland & IE & 80.65 & 52.38  \\
Belgium & BE & 70.27 & 45.82  &  Italy & IT & 79.41 & 44.55  \\
Bulgaria & BG & 72.97 & 37.88  &  Latvia & LV & 72.53 & 33.67  \\
Croatia & HR & 80.00 & 38.36  &  Lithuania & LT & 75.00 & 41.78  \\
Cyprus & CY & 79.17 & 64.95  &  Luxembourg & LU & 72.22 & 44.60  \\
Czech Republic & CZ & 71.70 & 35.98  &  Malta & MT & 80.56 & 66.37  \\
Denmark & DK & 77.50 & 54.09  &  Netherlands & NL & 74.55 & 48.18  \\
Estonia & EE & 66.67 & 36.46  &  Poland & PL & 75.00 & 31.62  \\
Finland & FI & 70.97 & 40.04  &  Portugal & PT & 81.54 & 51.33  \\
France & FR & 65.79 & 37.37  &  Romania & RO & 75.76 & 38.06  \\
Germany & DE & 67.57 & 30.24  &  Spain & ES & 74.07 & 43.06  \\
Great Britain & GB & 75.00 & 50.28  &  Slovakia & SK & 70.37 & 35.00  \\
Greece & GR & 77.19 & 40.94  &  Slovenia & SI & 78.00 & 37.78  \\
Hungary & HU & 75.44 & 43.80  &  Sweden & SE & 73.97 & 54.53  \\
& & & & European Union & EU & 73.25 & 40.63 \\
   \hline
\end{tabular}
\end{adjustbox}
\caption{Percentage of EU FB users (FFB) and citizens (FC) per EU Country that has been assigned some of the Top 500 sensitive ad preferences within their country. The last row reports the aggregated number of all 28 EU countries together.}
\label{table:top_interests}
\end{table}

\begin{table*}[t]
\centering
\resizebox{\hsize}{!}{
\begin{tabular}{lrrrrrrrrrrrrrrrrrrrrrrrrrrrrr}
    \hline
name & AT & BE & BG & HR & CY & CZ & DK & EE & FI & FR & DE & GR & HU & IE & IT & LV & LT & LU & MT & NL & PL & PT & RO & SK & SI & ES & SE & GB & EU28 \\ 
  \hline
COMMUNISM & 0.48 & 0.61 & 1.35 & 1.30 & 1.67 & 3.21 & 0.38 & 0.61 & 0.52 & 2.29 & 0.43 & 0.81 & 0.74 & 0.52 & 1.15 & 0.56 & 0.94 & 0.64 & 0.39 & 0.24 & 2.19 & 0.94 & 1.90 & 1.74 & 1.70 & 0.56 & 0.30 & 0.41 & 0.93\\ 
  ISLAM & 8.18 & 7.16 & 4.59 & 5.50 & 13.54 & 4.91 & 6.75 & 2.22 & 4.19 & 7.89 & 7.57 & 4.21 & 2.28 & 4.19 & 4.12 & 2.75 & 2.38 & 5.00 & 6.67 & 5.36 & 2.44 & 3.69 & 3.50 & 3.11 & 6.50 & 4.07 & 6.58 & 6.82 & 5.71 \\ 
  QURAN & 3.41 & 3.38 & 1.08 & 1.00 & 4.48 & 0.45 & 1.90 & 0.65 & 1.16 & 3.95 & 3.24 & 1.18 & 0.74 & 1.35 & 1.71 & 1.01 & 0.51 & 1.83 & 1.86 & 2.45 & 0.45 & 0.62 & 0.77 & 0.56 & 2.00 & 0.96 & 2.74 & 3.64 & 2.46 \\ 
  SUICIDE PREVENTION & 0.14 & 0.15 & 0.20 & 0.32 & 0.21 & 0.12 & 0.12 & 0.10 & 0.09 & 0.16 & 0.14 & 0.23 & 0.12 & 1.10 & 0.28 & 0.13 & 0.15 & 0.28 & 0.27 & 0.15 & 0.14 & 0.22 & 0.13 & 0.44 & 0.26 & 0.44 & 0.15 & 0.27 & 0.28\\ 
  SOCIALISM & 1.00 & 0.78 & 0.57 & 0.48 & 1.15 & 2.45 & 3.00 & 0.76 & 0.48 & 0.47 & 0.43 & 0.91 & 1.93 & 1.10 & 3.53 & 0.34 & 0.94 & 2.78 & 1.08 & 0.28 & 0.50 & 2.15 & 0.35 & 2.33 & 0.82 & 1.48 & 1.37 & 0.93  & 1.21\\ 
  JUDAISM & 2.50 & 1.16 & 0.86 & 0.70 & 2.29 & 0.72 & 2.17 & 1.01 & 0.61 & 1.26 & 1.38 & 1.30 & 1.16 & 1.26 & 2.29 & 1.76 & 1.81 & 1.19 & 3.06 & 1.00 & 1.19 & 1.69 & 1.40 & 0.93 & 0.74 & 1.15 & 0.64 & 0.95 & 1.32 \\ 
  HOMOSEXUALITY & 6.14 & 5.54 & 2.97 & 6.50 & 4.38 & 5.47 & 5.00 & 3.89 & 5.16 & 7.37 & 5.68 & 5.09 & 4.21 & 9.03 & 7.65 & 4.62 & 3.19 & 5.00 & 7.50 & 6.18 & 3.56 & 4.46 & 3.80 & 4.44 & 7.60 & 8.15 & 4.93 & 8.64 & 6.79\\ 
  ALTERNATIVE MEDICINE & 5.00 & 2.97 & 8.38 & 6.00 & 5.62 & 4.15 & 4.00 & 4.17 & 4.19 & 2.89 & 3.24 & 7.19 & 4.21 & 9.68 & 6.18 & 3.96 & 2.56 & 5.56 & 7.50 & 3.64 & 2.25 & 8.00 & 3.90 & 2.93 & 5.00 & 5.56 & 3.84 & 6.14 & 4.29\\ 
  CHRISTIANITY & 10.68 & 7.43 & 6.22 & 7.50 & 9.69 & 3.77 & 15.00 & 2.22 & 4.19 & 5.53 & 6.49 & 6.67 & 9.30 & 10.97 & 12.65 & 3.19 & 3.81 & 7.22 & 18.89 & 5.18 & 6.25 & 12.46 & 10.00 & 4.81 & 4.60 & 10.00 & 4.66 & 7.50 & 8.21\\ 
  ILLEGAL IMMIGRATION & 0.17 & 0.07 & 0.10 & 0.02 & 0.07 & 0.68 & 0.05 & 0.01 & 0.07 & 0.05 & 0.06 & 0.26 & 0.26 & 0.06 & 0.08 & 0.02 & 0.06 & 0.01 & 0.08 & 0.02 & 0.02 & 0.02 & 0.02 & 0.11 & 0.36 & 0.14 & 0.33 & 0.05 & 0.09\\ 
  ONCOLOGY & 0.23 & 0.27 & 0.62 & 0.44 & 3.96 & 0.57 & 0.15 & 0.10 & 0.08 & 0.17 & 0.16 & 0.49 & 0.30 & 1.29 & 0.94 & 0.70 & 1.62 & 0.19 & 0.78 & 0.45 & 1.25 & 1.09 & 0.73 & 0.59 & 0.21 & 0.70 & 0.08 & 0.66 & 0.61\\ 
  LGBT COMMUNITY & 6.36 & 6.62 & 5.14 & 6.50 & 6.56 & 6.04 & 6.50 & 5.14 & 6.45 & 7.11 & 5.95 & 5.79 & 4.39 & 11.94 & 8.53 & 5.27 & 5.88 & 6.67 & 9.44 & 6.36 & 5.88 & 7.85 & 6.30 & 4.81 & 6.00 & 7.04 & 6.44 & 11.14 & 8.21\\ 
  GENDER IDENTITY & 0.03 & 0.08 & 0.01 & 0.08 & 0.88 & 0.02 & 0.03 & 0.02 & 0.02 & 0.07 & 0.03 & 0.56 & 0.07 & 0.23 & 0.07 & 0.20 & 0.10 & 0.10 & 0.14 & 0.03 & 0.05 & 0.05 & 0.04 & 0.01 & 0.08 & 0.07 & 0.09 & 0.55 & 0.10\\ 
  REPRODUCTIVE HEALTH & 0.01 & 0.07 & 0.20 & 0.40 & 0.02 & 0.14 & 0.05 & 0.02 & 0.06 & 0.01 & 0.01 & 0.04 & 0.10 & 0.71 & 0.04 & 0.07 & 0.05 & 0.01 & 0.24 & 0.03 & 0.01 & 0.04 & 0.01 & 0.03 & 0.00 & 0.03 & 0.05 & 0.13 & 0.07\\ 
  BIBLE & 17.95 & 10.81 & 8.65 & 10.50 & 11.46 & 7.17 & 12.75 & 4.31 & 4.84 & 7.63 & 15.41 & 8.25 & 10.00 & 19.03 & 17.65 & 5.71 & 6.25 & 14.44 & 20.28 & 10.91 & 14.38 & 12.31 & 8.70 & 6.67 & 7.40 & 7.04 & 5.48 & 15.68 & 12.14\\ 
  PREGNANCY & 15.68 & 12.97 & 9.19 & 17.00 & 13.54 & 16.23 & 14.50 & 10.00 & 11.29 & 10.79 & 11.89 & 13.51 & 11.23 & 20.97 & 12.35 & 13.19 & 18.75 & 12.78 & 9.72 & 14.55 & 15.00 & 18.46 & 9.70 & 18.89 & 13.00 & 14.07 & 13.42 & 18.41 & 14.29 \\ 
  NATIONALISM & 0.86 & 0.78 & 1.65 & 1.85 & 2.19 & 2.45 & 1.00 & 0.58 & 0.45 & 1.08 & 1.00 & 1.74 & 2.11 & 2.00 & 1.32 & 2.42 & 0.94 & 2.19 & 2.78 & 0.70 & 3.00 & 1.69 & 2.50 & 1.37 & 0.61 & 1.11 & 0.99 & 0.91 & 1.39\\ 
  VEGANISM & 14.55 & 10.27 & 7.30 & 10.50 & 10.21 & 9.25 & 12.75 & 9.86 & 15.16 & 8.68 & 11.35 & 9.82 & 9.82 & 14.84 & 13.53 & 9.23 & 8.12 & 13.06 & 13.33 & 10.91 & 8.12 & 11.23 & 6.70 & 8.52 & 14.00 & 10.37 & 16.44 & 13.64 & 11.43\\ 
  BUDDHISM & 3.18 & 3.38 & 1.62 & 3.55 & 3.33 & 2.26 & 2.08 & 1.53 & 1.13 & 2.61 & 1.43 & 2.63 & 3.33 & 3.87 & 2.94 & 1.98 & 1.88 & 3.33 & 4.17 & 2.45 & 1.31 & 6.92 & 1.90 & 1.67 & 3.00 & 2.19 & 1.51 & 2.50 & 2.39\\ 
  FEMINISM & 4.55 & 3.78 & 3.51 & 3.80 & 5.52 & 2.08 & 5.50 & 2.78 & 6.77 & 5.00 & 3.78 & 3.68 & 2.46 & 9.35 & 5.88 & 3.19 & 3.56 & 5.83 & 8.61 & 3.64 & 3.44 & 8.15 & 2.40 & 4.07 & 3.90 & 8.89 & 13.70 & 7.27 & 7.50 \\ 
  UNION & 45.45 & 39.19 & 32.43 & 41.50 & 45.83 & 37.74 & 45.00 & 27.78 & 35.48 & 34.21 & 40.54 & 36.84 & 36.84 & 51.61 & 44.12 & 32.97 & 36.25 & 41.67 & 47.22 & 40.00 & 36.88 & 44.62 & 34.34 & 35.56 & 39.00 & 40.74 & 41.10 & 47.73 & 42.86\\
   \hline
\end{tabular}}
\caption{Percentage of FB users (FFB) per EU country that have been assigned each of the 20 very sensitive ad preferences listed in the table. The last row reports the aggregated FFB value for all 20 ad preferences per EU country. The last column reports the aggregated FFB value across all 28 EU countries.}
\label{table:sens1}
\end{table*}

\begin{table*}[t]
\centering
\resizebox{\hsize}{!}{
\begin{tabular}{lrrrrrrrrrrrrrrrrrrrrrrrrrrrrr}
    \hline
name & AT & BE & BG & HR & CY & CZ & DK & EE & FI & FR & DE & GR & HU & IE & IT & LV & LT & LU & MT & NL & PL & PT & RO & SK & SI & ES & SE & GB & EU28 \\ 
  \hline
COMMUNISM & 0.24 & 0.40 & 0.70 & 0.62 & 1.37 & 1.61 & 0.26 & 0.33 & 0.29 & 1.30 & 0.19 & 0.43 & 0.43 & 0.34 & 0.64 & 0.26 & 0.52 & 0.39 & 0.32 & 0.15 & 0.92 & 0.59 & 0.96 & 0.87 & 0.82 & 0.32 & 0.22 & 0.27 & 0.51\\ 
  ISLAM & 4.12 & 4.67 & 2.39 & 2.64 & 11.11 & 2.46 & 4.71 & 1.22 & 2.37 & 4.48 & 3.39 & 2.23 & 1.32 & 2.72 & 2.31 & 1.28 & 1.32 & 3.09 & 5.49 & 3.47 & 1.03 & 2.32 & 1.78 & 1.55 & 3.15 & 2.37 & 4.85 & 4.57 & 3.13\\ 
  QURAN & 1.71 & 2.20 & 0.56 & 0.48 & 3.67 & 0.23 & 1.33 & 0.36 & 0.66 & 2.24 & 1.45 & 0.62 & 0.43 & 0.88 & 0.96 & 0.47 & 0.28 & 1.13 & 1.53 & 1.59 & 0.19 & 0.39 & 0.39 & 0.28 & 0.97 & 0.56 & 2.02 & 2.44 & 1.35\\ 
  SUICIDE PREVENTION & 0.07 & 0.10 & 0.10 & 0.15 & 0.17 & 0.06 & 0.08 & 0.05 & 0.05 & 0.09 & 0.06 & 0.12 & 0.07 & 0.71 & 0.16 & 0.06 & 0.08 & 0.17 & 0.22 & 0.09 & 0.06 & 0.14 & 0.07 & 0.22 & 0.13 & 0.26 & 0.11 & 0.18 & 0.15\\ 
  SOCIALISM & 0.50 & 0.51 & 0.29 & 0.23 & 0.94 & 1.23 & 2.09 & 0.42 & 0.27 & 0.27 & 0.19 & 0.48 & 1.12 & 0.71 & 1.98 & 0.16 & 0.52 & 1.72 & 0.89 & 0.18 & 0.21 & 1.36 & 0.18 & 1.16 & 0.40 & 0.86 & 1.01 & 0.62 & 0.66\\ 
  JUDAISM & 1.26 & 0.76 & 0.45 & 0.34 & 1.88 & 0.36 & 1.52 & 0.55 & 0.35 & 0.72 & 0.62 & 0.69 & 0.67 & 0.82 & 1.29 & 0.82 & 1.01 & 0.74 & 2.52 & 0.65 & 0.50 & 1.07 & 0.71 & 0.46 & 0.36 & 0.67 & 0.47 & 0.64 & 0.72\\ 
  HOMOSEXUALITY & 3.09 & 3.61 & 1.54 & 3.12 & 3.59 & 2.75 & 3.49 & 2.13 & 2.91 & 4.19 & 2.54 & 2.70 & 2.44 & 5.87 & 4.29 & 2.14 & 1.78 & 3.09 & 6.18 & 4.00 & 1.50 & 2.81 & 1.93 & 2.21 & 3.68 & 4.74 & 3.64 & 5.79 & 3.71\\ 
  ALTERNATIVE MEDICINE & 2.52 & 1.94 & 4.35 & 2.88 & 4.61 & 2.08 & 2.79 & 2.28 & 2.37 & 1.64 & 1.45 & 3.82 & 2.44 & 6.29 & 3.47 & 1.84 & 1.43 & 3.43 & 6.18 & 2.35 & 0.95 & 5.04 & 1.98 & 1.46 & 2.42 & 3.23 & 2.83 & 4.11 & 2.34\\ 
  CHRISTIANITY & 5.37 & 4.85 & 3.23 & 3.60 & 7.95 & 1.89 & 10.47 & 1.22 & 2.37 & 3.14 & 2.90 & 3.54 & 5.40 & 7.12 & 7.10 & 1.48 & 2.12 & 4.46 & 15.56 & 3.35 & 2.64 & 7.85 & 5.07 & 2.39 & 2.23 & 5.81 & 3.43 & 5.03 & 4.49\\ 
  ILLEGAL IMMIGRATION & 0.09 & 0.04 & 0.05 & 0.01 & 0.06 & 0.34 & 0.03 & 0.00 & 0.04 & 0.03 & 0.03 & 0.14 & 0.15 & 0.04 & 0.04 & 0.01 & 0.03 & 0.01 & 0.07 & 0.01 & 0.01 & 0.01 & 0.01 & 0.05 & 0.17 & 0.08 & 0.24 & 0.04 & 0.05\\ 
  ONCOLOGY & 0.11 & 0.18 & 0.32 & 0.21 & 3.25 & 0.28 & 0.10 & 0.06 & 0.05 & 0.10 & 0.07 & 0.26 & 0.17 & 0.84 & 0.53 & 0.33 & 0.91 & 0.12 & 0.64 & 0.29 & 0.53 & 0.69 & 0.37 & 0.29 & 0.10 & 0.41 & 0.06 & 0.44 & 0.33\\ 
  LGBT COMMUNITY & 3.20 & 4.32 & 2.67 & 3.12 & 5.38 & 3.03 & 4.54 & 2.81 & 3.64 & 4.04 & 2.66 & 3.07 & 2.55 & 7.75 & 4.79 & 2.45 & 3.27 & 4.12 & 7.78 & 4.11 & 2.48 & 4.94 & 3.20 & 2.39 & 2.91 & 4.09 & 4.75 & 7.47 & 4.49\\ 
  GENDER IDENTITY & 0.01 & 0.05 & 0.01 & 0.04 & 0.72 & 0.01 & 0.02 & 0.01 & 0.01 & 0.04 & 0.01 & 0.30 & 0.04 & 0.15 & 0.04 & 0.09 & 0.06 & 0.06 & 0.12 & 0.02 & 0.02 & 0.03 & 0.02 & 0.00 & 0.04 & 0.04 & 0.06 & 0.37 & 0.05\\ 
  REPRODUCTIVE HEALTH & 0.00 & 0.05 & 0.11 & 0.19 & 0.02 & 0.07 & 0.04 & 0.01 & 0.04 & 0.01 & 0.00 & 0.02 & 0.06 & 0.46 & 0.02 & 0.03 & 0.03 & 0.01 & 0.19 & 0.02 & 0.01 & 0.02 & 0.01 & 0.01 & 0.00 & 0.02 & 0.03 & 0.09 & 0.04\\ 
  BIBLE & 9.03 & 7.05 & 4.49 & 5.04 & 9.40 & 3.60 & 8.90 & 2.35 & 2.73 & 4.34 & 6.90 & 4.37 & 5.81 & 12.36 & 9.90 & 2.65 & 3.48 & 8.92 & 16.71 & 7.05 & 6.06 & 7.75 & 4.42 & 3.32 & 3.58 & 4.09 & 4.04 & 10.51 & 6.64\\ 
  PREGNANCY & 7.89 & 8.46 & 4.77 & 8.15 & 11.11 & 8.14 & 10.12 & 5.47 & 6.37 & 6.13 & 5.32 & 7.16 & 6.52 & 13.62 & 6.93 & 6.12 & 10.44 & 7.89 & 8.01 & 9.40 & 6.32 & 11.62 & 4.92 & 9.39 & 6.30 & 8.18 & 9.90 & 12.34 & 7.82\\ 
  NATIONALISM & 0.43 & 0.51 & 0.86 & 0.89 & 1.79 & 1.23 & 0.70 & 0.32 & 0.25 & 0.61 & 0.45 & 0.92 & 1.22 & 1.30 & 0.74 & 1.12 & 0.52 & 1.36 & 2.29 & 0.45 & 1.26 & 1.07 & 1.27 & 0.68 & 0.30 & 0.65 & 0.73 & 0.61 & 0.76\\ 
  VEGANISM & 7.32 & 6.70 & 3.79 & 5.04 & 8.38 & 4.64 & 8.90 & 5.39 & 8.55 & 4.93 & 5.08 & 5.21 & 5.70 & 9.64 & 7.59 & 4.28 & 4.53 & 8.06 & 10.99 & 7.05 & 3.43 & 7.07 & 3.40 & 4.24 & 6.78 & 6.03 & 12.12 & 9.14 & 6.25\\ 
  BUDDHISM & 1.60 & 2.20 & 0.84 & 1.70 & 2.73 & 1.14 & 1.45 & 0.84 & 0.64 & 1.48 & 0.64 & 1.40 & 1.94 & 2.51 & 1.65 & 0.92 & 1.04 & 2.06 & 3.43 & 1.59 & 0.55 & 4.36 & 0.96 & 0.83 & 1.45 & 1.27 & 1.11 & 1.68 & 1.31\\ 
  FEMINISM & 2.29 & 2.47 & 1.82 & 1.82 & 4.53 & 1.04 & 3.84 & 1.52 & 3.82 & 2.84 & 1.69 & 1.95 & 1.43 & 6.08 & 3.30 & 1.48 & 1.98 & 3.60 & 7.09 & 2.35 & 1.45 & 5.13 & 1.22 & 2.03 & 1.89 & 5.17 & 10.10 & 4.88 & 4.10\\ 
  UNION & 22.86 & 25.55 & 16.84 & 19.90 & 37.60 & 18.94 & 31.41 & 15.19 & 20.02 & 19.43 & 18.14 & 19.54 & 21.39 & 33.52 & 24.75 & 15.30 & 20.19 & 25.73 & 38.91 & 25.85 & 15.55 & 28.09 & 17.25 & 17.68 & 18.89 & 23.68 & 30.29 & 31.99 & 23.45\\ 
   \hline
\end{tabular}}
\caption{Percentage of citizens (FC) per EU country that have been assigned each of the 20 very sensitive ad preferences listed in the table. The last row reports the aggregated FC value for all 20 ad preferences per EU country. The last column reports the aggregated FC value across all 28 EU countries.}
\label{table:sens2}
\end{table*}

\subsection{Very Sensitive Ad Preferences}

The GDPR clearly defines what \emph{sensitive data} is. Online services should guarantee that sensitive data is neither collected nor processed to comply with the legislation unless some of the exceptions apply (see Section \nameref{sec:legal}). Despite this consideration, most people probably agree that all sensitive information is not equally sensitive. For instance, data revealing the sexual preference of a user is in principle more sensitive than data revealing a user may be on a diet to lose some weight.

We acknowledge that the level of sensitivity varies across sensitive ad preferences included in our dataset. Hence, although they have been classified as sensitive by 12 panelists, someone may still argue that in his or her opinion some of the sensitive ad preferences in our dataset are not actually sensitive and thus the results reported in the previous subsection may not be accurate. 

In this subsection, we prove that even in the case when we focus on manually selected sensitive ad preferences, which doubtlessly matches the GDPR definition of \textit{sensitive data}, the percentage of EU FB users and citizens affected is still very high.

In particular, we have selected 20 ad preferences that we refer to as \emph{very} sensitive. A \emph{very} sensitive ad preference is defined as an ad preference that all the 12 panelists and the authors would consider offensive if it were assigned to him/herself or a relative (e.g., spouse, children, parents). Tables \ref{table:sens1} and \ref{table:sens2} show the percentage of FB users (FFB) and citizens (FC) tagged with each of the 20 \emph{very} sensitive ad preferences per EU country. Note that the last row presents the aggregate results for the 20 very sensitive ad preferences in each country, and the last column presents the aggregate results for the 28 EU countries together.

We observe that 42.9\% EU FB users, which corresponds to 23.5\% EU citizens, are tagged with at least one of the \emph{very} sensitive ad preferences, respectively. Hence, around one-quarter of the EU population has been tagged in FB with ad preferences that are doubtlessly related to \textit{sensitive data} as reported by the GDPR. If we analyze the results per country, we observe that the fraction of the population affected ranges between 15\% in Estonia (EE), Latvia (LV) and Poland (PL) and 38\% in Malta (MT). These observations confirm the existence of a worrisome privacy problem for EU citizens. 


\subsection{Age and Gender analysis}

We analyze the exposition of different demographic groups (based on gender and age) to sensitive ad preferences. The gender analysis considers two groups, men vs. women, while the age analysis considers four age groups following the division proposed by Erikson et al. \cite{erikson1998life}: 13-19 (Adolescence), 20-39 (Early Adulthood), 40-64 (Adulthood) and 65+ (Maturity). For each group, we compute FFB(C = EU28, N = 500) and FFB(C = EU28, N = 20 very sensitive ad preferences). Figures \ref{fig:age_groups} and \ref{fig:genders_groups} report the results for age and gender groups, respectively. 


The Early Adulthood is clearly the most exposed group to sensitive ad preferences. Especially, 61\% (45\%) users in this group have been tagged with some of the Top 500 (20-very) sensitive ad preferences. Following the Early Adulthood group we find the Adolescence, Adulthood and Maturity groups with 55\% (42\%), 40\% (32\%) and 39\% (28\%) of its users tagged with some of the Top 500 (20-very) sensitive ad preferences, respectively. Although the difference in the exposition to sensitive ad preferences is substantial across groups, all of them present a considerably high exposure. In particular, more than one-quarter of the users within every group is exposed to very sensitive ad preferences.

The gender-based analysis shows that women exposition to the Top 500 (20-very) sensitive ad preferences reaches 78\% (49\%). The exposition is notably smaller for men, where the fraction of tagged users with some of the Top 500 (20-very) sensitive ad preferences shrinks by 10 (18) percentage points to 68\% (31\%). This result suggests the existence of a gender bias, which despite its obvious interest is out of the scope of this paper.

\section{Proof of commercial exploitation of sensitive ad preferences with real FB Ad Campaigns}

\label{subsec:ads_campaigns}

\begin{figure}[t]
 	\centering
 	\includegraphics[width=1\columnwidth]{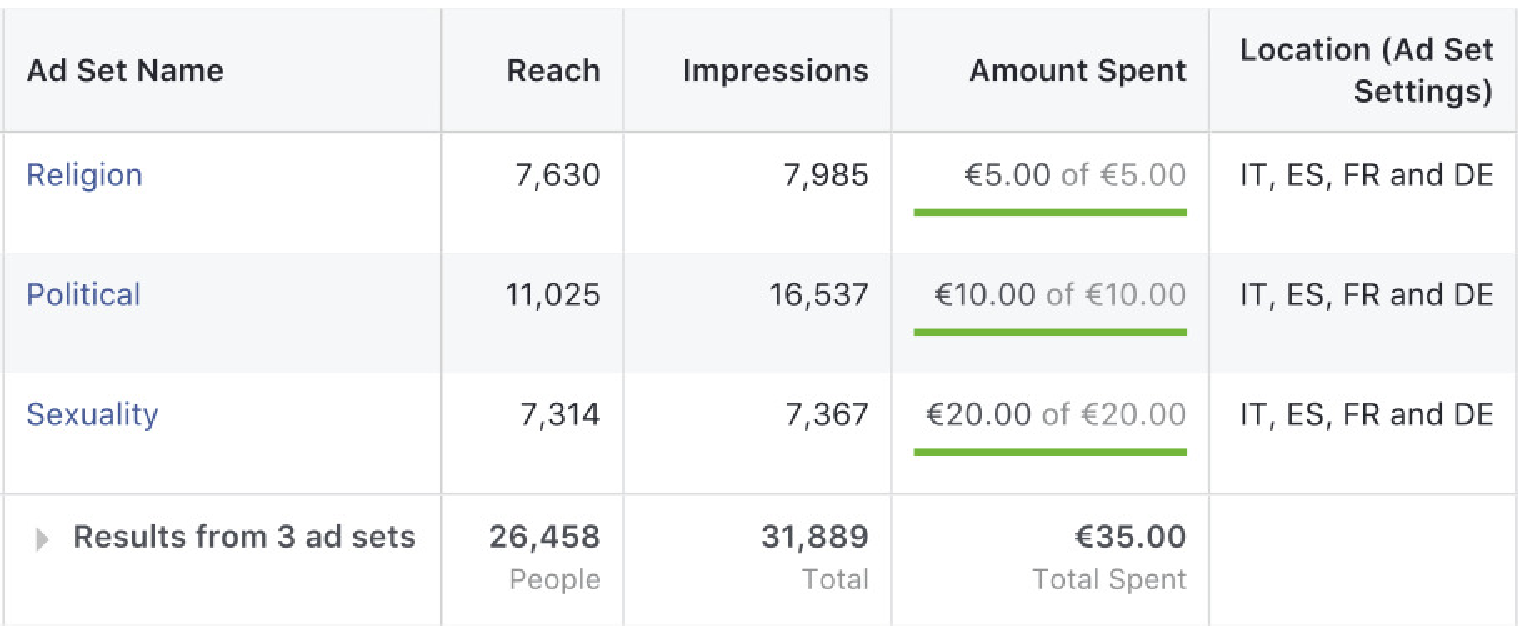}\hfill
 	\caption{FB report from the 3 ad campaigns we ran targeting users based on sensitive ad preferences.}
 	\label{fig:FB_campaigns}
\end{figure}

The analysis conducted so far has just proved that Facebook labels a significant portion of EU citizens using sensitive personal data. In this section, we prove that FB is exploiting sensitive personal data for commercial purposes through advertising campaigns. We ran 3 FB ad campaigns using sensitive ad preferences: \textit{``religious beliefs"} (targeting users interested in Islam OR Judaism OR Christianity OR Buddhism), \textit{``political opinions"} (targeting users interested in Communism OR Anarchism OR Radical feminism OR Socialism) and \textit{``sexual orientation"} (targeting users interested in Transsexualism OR Homosexuality). The 3 campaigns focused on four EU countries: Germany, Spain, France and Italy. 

Overall, with a budget of \euro35 we were able to reach 26458 users tagged with some of the previous sensitive ad preferences. Our credit card was charged and we received the bills associated with our campaigns. Figure \ref{fig:FB_campaigns} shows a snapshot of the Facebook summary report for each campaign. This experiment provides substantial evidence that Facebook generates revenue from the commercial exploitation of sensitive personal data according to the GDPR definition of \textit{sensitive data}. 

Note that the conducted ad campaigns were compliant with the Terms \& Services of Facebook introduced in Section \nameref{subsec:fb_ts} 

\begin{figure}[ht]
	\centering
	\begin{minipage}[t]{0.48\hsize}
		\centering
		\includegraphics[width=1\textwidth]{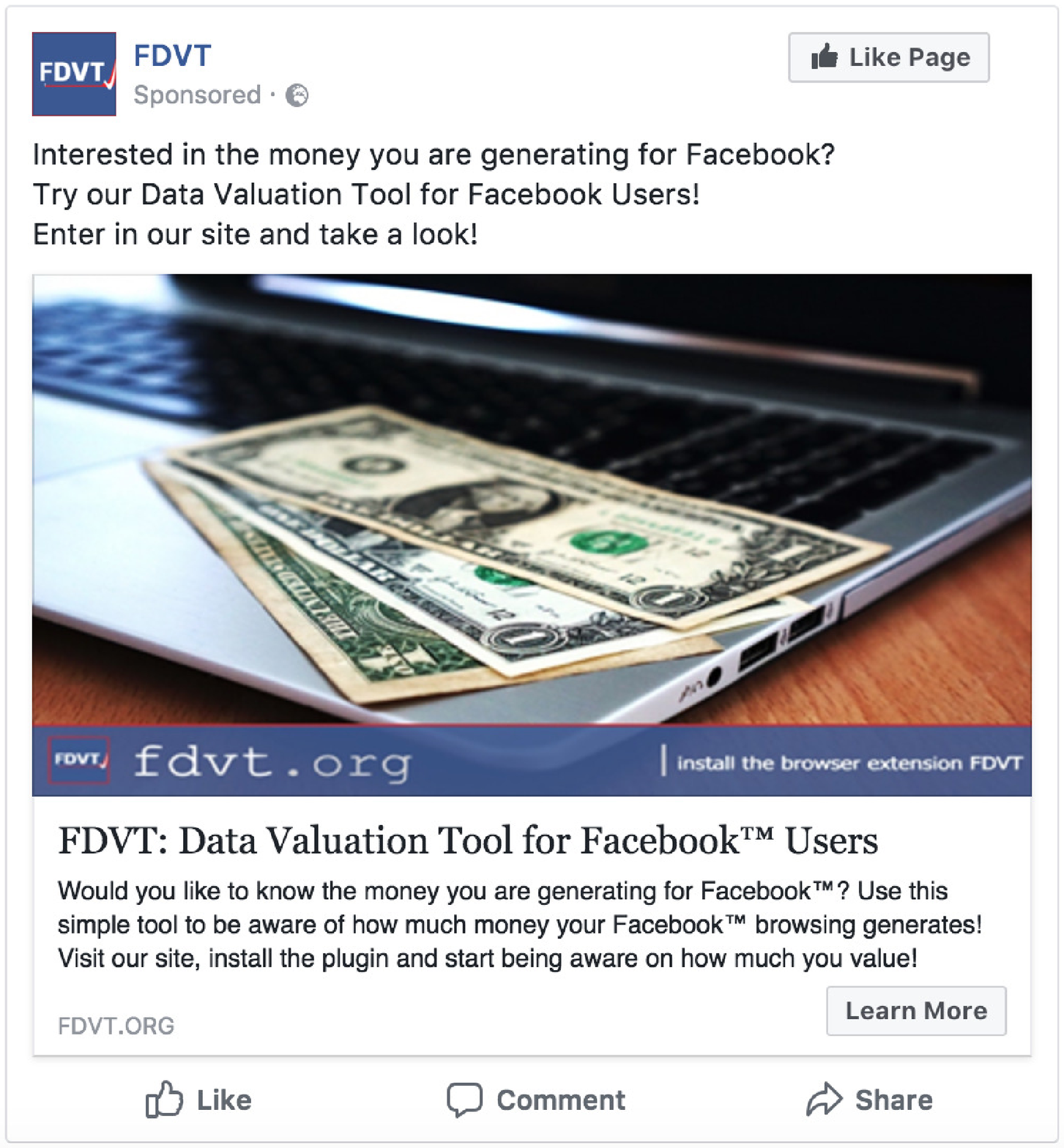}\hfill
		\caption{FDVT ad 1}
		\label{fig:FDVT_ad1}
	\end{minipage}
	\hfill
	\begin{minipage}[t]{0.48\hsize}
		\centering
		\includegraphics[width=1\textwidth]{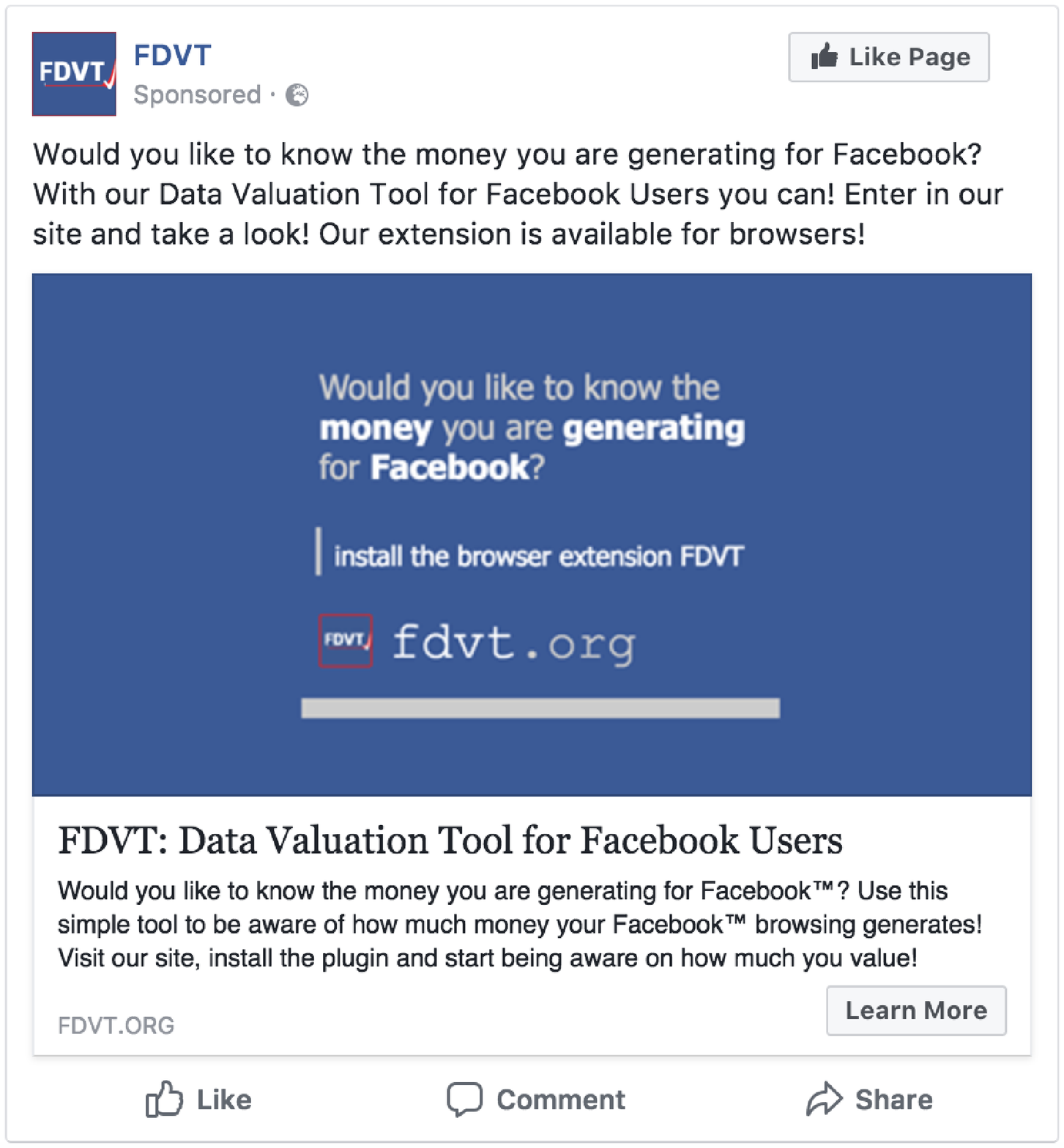}\hfill
		\caption{FDVT ad 2}
		\label{fig:FDVT_ad2}
	\end{minipage}
\end{figure}
Figures \ref{fig:FDVT_ad1} and \ref{fig:FDVT_ad2} show the two ads we used in our campaigns. These ads refer to our FDVT browser extension and thus they do not include content that asserts or implies personal attributes. Indeed, the landing page where users were redirected in case they clicked in any of these ads is the webpage of the FDVT project\footnote{\url{https://www.fdvt.org/}}.

Moreover, our experiments were compliant with the data protection legislation applicable to the authors' institution. In particular, in the experiments, we did not record any information from those users clicking the ads and visiting our landing page. The only information we use in this paper is the one provided by FB through the reports it offers to advertisers related to their ad campaigns.

\section{Ethics and Privacy Risks Associated with Sensitive Personal Data Exploitation}
\label{sec:discussion}

The possibility of reaching users labeled with sensitive personal data enables the use of FB ads campaigns to attack specific groups of people based on sensitive personal data (race, sexual orientation, religious beliefs, etc.). Following, we illustrate two specific examples of potential attacks.

\noindent {\textbf{Hate campaigns}:} An attacker could create hate-of-speech campaigns using sensitive ad preferences representative of a specific sensitive social group within its target audience. For instance, a neo-nazi organization could create ads campaigns with offensive messages targeting people interested in \textit{Judaism} or \textit{Homosexuality}. This clearly presents important ethical concerns since such hate-of-speech campaigns can reach thousands of users at a very low cost (e.g., we reached more than 26K FB users spending only \euro35 in FB ads campaigns).


\noindent \textbf{Identification attack:} An attacker can use FB as a proxy to identify citizens belonging to a sensitive social group defined by its religious belief, sexual orientation, political preference, etc. To this end, an attacker just needs to replicate a phishing-like attack \cite{phising_survey}. The attacker would configure a campaign targeting a sensitive audience (e.g., people interested in \textit{homosexuality}) using a fancy advertisement that serves as bait to attract the targeted users to the attacker's webpage (e.g., the ad promises the user to win an iPhone X if she clicks on the ad). If the user clicks on the ad, she will be redirected to the attacker's webpage. Once there, the attacker can use different techniques exploited in phishing attacks \cite{phising_survey} persuading the user to provide some personal data that would reveal her identity.  For instance, in the example of the iPhone X, the landing page can show a message congratulating the user for winning the phone, and at the same time, it may request the user to provide some personal data (name, address, phone number, etc.) for shipping purposes.

A recent study \cite{phiseye} ran experiments implementing email-based phishing attacks in which 9\% of the users posted their credentials (username and password) in the phishing site (i.e., attacker's landing page). 
Using as a reference this success rate for phishing attacks and the results from the ad campaigns described in Section \nameref{subsec:ads_campaigns}, we can make a ball-park estimation of the cost of identifying users tagged with sensitive ad preferences. We spent \euro35 in our ads campaigns to reach 26K users from which 2.34K (according to the 9\% reference success rate) may fill some personal information in the attacker's webpage that would unveil their identity. Based on this, identifying an individual user may be as cheap as \euro0.015. Even if we consider a success rate two orders of magnitude smaller (0.09\%), the cost would be \euro1.5 per user.

The estimated cost to unveil the identity of users based on sensitive personal data is rather low considering the serious privacy risks users may face. For instance, $(i)$ in countries where homosexuality is considered illegal or immoral governments or other organizations could obtain the identity of people that are likely homosexual (e.g., interested in \textit{homosexuality, LGBT}, etc.); $(ii)$ neo-nazis organizations could identify people in specific regions (using as location target a town or even a zip code) that are likely Jewish (e.g., interested in \textit{Judaism, Shabbat}, etc.); $(iii)$ health assurance companies could try to identify people that may have non-profitable habits (e.g., interested in \textit{tobacco, fast food}, etc.) or health problems (e.g., \textit{food intolerance}) for not accepting them as clients or increasing the health insurance cost for them. Besides, users may face the negative consequences of such phishing-like attacks even in the case that FB has wrongly labeled them with some sensitive ad preference.

In summary, although Facebook does not allow third parties directly accessing individual users' identity, ad preferences can be used as a very powerful proxy to perform identification attacks\footnote{The described attack can be implemented on any advertising platform allowing advertisers to target users based on sensitive personal data.} based on sensitive personal data at a low cost. Note that we have simply illustrated the ad-based phishing attack but not implemented it due to the obvious ethical implications.

\section{FDVT extension to inform users about their sensitive ad preferences}
\label{sec:fdvt_extension}

The results reported in previous sections urge to provide solutions that make users aware of the use of sensitive personal data for advertising purposes. To this end, we have extended the FDVT browser extension to inform users about the sensitive ad preferences that FB has assigned them: $(i)$ we have built a classifier to automatically tag ad preferences assigned to FDVT users as sensitive or non-sensitive; $(ii)$ we have modified the FDVT back-end and front-end to incorporate this new feature.
\\
 
\begin{figure}
\centering
	\includegraphics[width=.75\columnwidth]{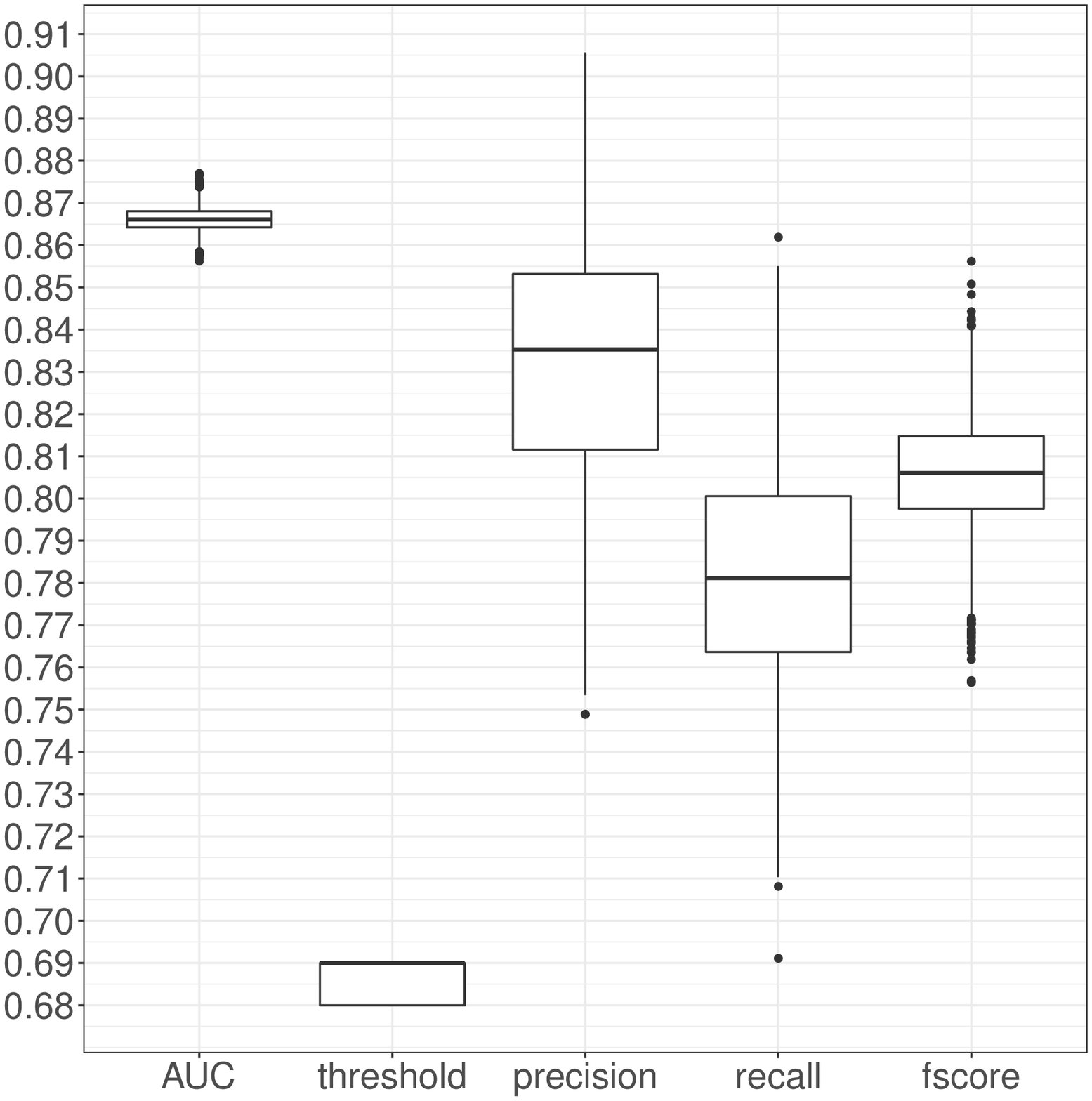}\hfill
	\caption{Test}
	\label{fig:test_training}
\end{figure}
\vspace{-1.5em}
 
\subsection{Automatic Binary Classifier for Sensitive Ad Preferences}

We rely on the methodology described in Section \nameref{sec:methodology} to compute the semantic similarity between ad preferences and sensitive personal data categories (i.e., politics, religion, health, ethnicity and sexual orientation). We remind that each ad preference is assigned a semantic similarity score that ranges between 0 (lowest) and 1 (highest). To build an automatic binary classifier we have to define a threshold so that ad preferences over (below) it are classified as sensitive (non-sensitive). 



To set up this threshold, we use the manually labeled dataset from Section \nameref{subsec:manual_classification}. It includes 4452 ad preferences , where 2092 were classified as sensitive. 
We follow a standard training-testing model approach. We randomly split our dataset in a training and a validation subsets that include 80\% and 20\% of the samples, respectively. The training subset is used to find the optimal threshold. In turn, we use the validation subset to assess the performance of the selected threshold. The optimal threshold is selected as the one maximizing the F-score for the training subset \cite{Ricci:2010:RSH:1941884}. Moreover, we validate the performance of the selected threshold computing the precision, recall and F-score on the validation subset. We have performed 5000 realizations of this process, each using different randomly chosen testing and validation subsets, to prove the robustness of the proposed binary classifier.

Figure \ref{fig:test_training} presents boxplots showing the AUC, precision, recall and F-score for the optimal threshold across the 5000 realizations. The optimal threshold remains very stable ranging between 0.68 and 0.69. Similarly, the AUC derived from the ROC curve from our binary classifier presents a very stable result around 0.86, which is associated with a good performance according to standard quality metrics in research \cite{fawcett2006introduction}\cite{zhu2010sensitivity}. 

Finally, the median precision of our binary classifier is 0.835 (min = 0.75, max = 0.90) and the median recall is 0.78 (min = 0.70, max = 0.86).

In summary, we conclude that the performance of the proposed binary classifier is good enough to accomplish our goal of showing FDVT users which of their ad preferences may be linked to sensitive personal data. In turn, we expect that this information helps to create an increasing collective awareness among FB users regarding the use of sensitive personal data for advertising purposes.

\begin{figure}[t]
	\centering
	\includegraphics[width=1\columnwidth]{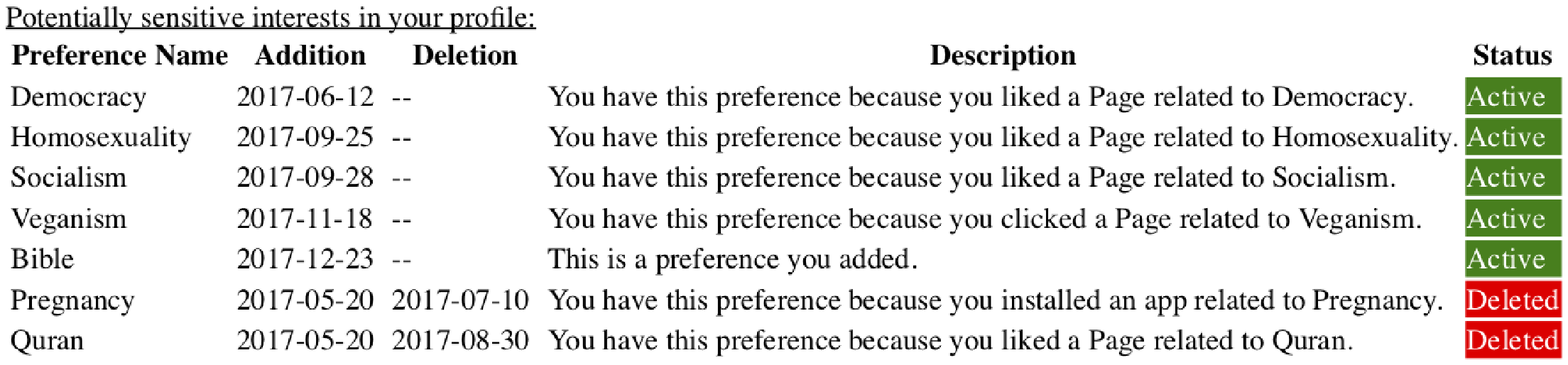}\hfill
	\caption{Webpage displaying sensitive preferences.}
	\label{fig:FDVT_example2}
\end{figure}

\subsection{System implementation}

\noindent \textbf{FDVT Backend:} We have computed the semantic similarity score for all ad preferences stored in our database. We have classified as sensitive and introduce in a blacklist those ad preferences with a similarity score $\geq$0.69.\footnote{The value of the optimal threshold may change over the time since it will be recomputed periodically.} Each time a FDVT user starts a session in FB we retrieve her updated set of ad preferences and compare them with the blacklist to obtain the list of ad preferences linked to sensitive personal data. We store the history of sensitive ad preferences assigned to the user to being able to notify her of sensitive ad preferences that FB has removed. Finally, every time a user is assigned a new ad preference that was not registered before in our database, we compute its semantic similarity score and include it in the blacklist if the ad preference is classified as sensitive.


\noindent \textbf{FDVT User Interface:} We have introduced a new button in the FDVT extension interface with the label \textit{"Sensitive FB Preferences"}. When a user clicks on that button, we display a web page listing the sensitive ad preferences included in the user's ad preference set. Figure \ref{fig:FDVT_example2} shows a snapshot of an example of such webpage. We provide the following information for each sensitive ad preference: $(i)$ Ad preference name, $(ii)$ Addition date, $(iii)$ Deletion date (only for removed sensitive ad preferences), $(iv)$ Description, which indicates the reason why FB has assigned that ad preference to the user, $(v)$ Status, either active (highlighted in green) or deleted (highlighted in red).




\section{Related Work}


We focus on discussing previous works that have addressed issues associated with sensitive personal data in online advertising and recent works that analyze privacy and discrimination issues related to FB advertising and ad preferences.

Carrascosa et al. \cite{Carrascosa} propose a new methodology to quantify the portion of targeted ads received by Internet users while they browse the web. They create bots, referred to as \textit{personas}, with very specific interests profiles (e.g., persona interested in cars) and measure how many of the received ads actually match the specific interest of the analyzed persona. They create personas based on sensitive personal data (e.g., health) and demonstrate that they are also targeted with ads related to the sensitive information used to create the persona's profile.      
Castellucia et al. \cite{castelluccia2012betrayed} shown that an attacker that gets access (e.g., through a public WiFi network) to the Google ads received by a user could create an interest profile that could reveal up to 58\% of the actual interests of the user. They state that in the case that some of the unveiled interests is sensitive, it could imply serious privacy risks for users. 

Venkatadri et al. have recently presented their work that exposes privacy \cite{EURECOM+5420} and discrimination \cite{speicherpotential} vulnerabilities related to FB advertising. In \cite{EURECOM+5420}, the authors demonstrate how an attacker can use Facebook third-party tracking javascript to retrieve personal data (e.g., mobile phone number) associated with users visiting the attacker's website. Moreover, in \cite{speicherpotential} they demonstrate that sensitive FB ad preferences can be used to apply negative discrimination in advertising campaigns (e.g., excluding people based on their race). They also show that some ad preferences that initially may not seem sensitive could be used to discriminate some social collective in advertising campaigns (e.g., excluding people interested in \textit{Blacknews.com} that are potentially black people). 

Finally, Andreou et al. \cite{andreou2018investigating} analyse whether the reasons FB uses to explain why a user is targeted with an ad are aligned with the actual audience the advertiser is targeting. To this, they analyze an explanation that Facebook includes in each delivered ad referred to as \textit{``Why Am I Seeing this Ad"}. This explanation describes the target audience associated with the delivered ad.  Out of the analysis of 79 ads, they conclude that in many cases the provided explanations are incomplete and sometimes misleading. They also perform a qualitative analysis related to the ad preferences assigned to FB user based on a small dataset including 9K ad preferences distributed across 35 users. They conclude that the reasons why ad preferences are assigned are vague.


In summary, the existing literature suggests that the online advertising ecosystem (beyond Facebook) exploits sensitive personal information for commercial purposes. In addition, there are previous works that highlight several privacy and ethics vulnerabilities associated with FB ad preferences. Our work completes  this body of literature quantifying the number of users in FB that are exposed to the commercial exploitation of their sensitive personal data.

\section{IRB and FDVT users' consent}
\label{sec:IRB}


The Ethics committee of the authors' institution has provided IRB approval to conduct the implementation of the FDVT and the research activities derived from it.

To comply with the most rigorous ethics and legal standards, during the installation process of the FDVT, a user has to: $(i)$ read and accept the Terms of Use\footnote{\url{https://www.fdvt.org/terms_of_use/}} and privacy policy\footnote{\url{https://www.fdvt.org/privacy_agreement.html}}; $(ii)$  grant explicit permission to use the information stored (in an anonymous manner) for research purposes.

Finally, it is also worth mentioning that we did not gather any information (neither personal or none personal) from those users who clicked in the ads we used in the FB advertising campaigns described in Section \nameref{subsec:ads_campaigns}.

\section{Conclusion}

Facebook is commercially exploiting sensitive personal data for advertising purposes through the so-called ad preferences that reveal potential interests of Facebook users. Facebook has already been fined in Spain for this practice. In May 2018 a new data protection regulation (GDPR), which prohibits the processing of sensitive personal data, will be enforced in all EU Countries. The short time until the GDPR application has been one of our main motivations to measure the portion of EU FB users, and the correspondent portion of citizens, that have been assigned ad preferences linked to sensitive personal data. The results reveal that such portion is as high as 73\% FB users (40\% EU citizens). We illustrate how FB users that have been assigned sensitive ad preferences could face serious privacy risks since the identity of some of them could be unveiled at low-cost through simple phishing-like attacks. The results of our paper urge a quick reaction from Facebook to eliminate from its ad preferences list all those that can be used to infer the politic orientation, sexual orientation, health conditions, religious believes or ethnic origin of a user for two reasons: $(i)$ this will guarantee that Facebook complies with the GDPR, $(ii)$ it will preserve the privacy of the users from attackers that aim to unveil the identity of groups of people linked to (very) sensitive information.


\small
\bibliographystyle{acm-sigchi}
\bibliography{references}

\end{document}